\documentclass[prd,aps,nofootinbib,nobibnotes,superscriptaddress,preprintnumbers,11pt]{revtex4}

\usepackage{color}%to be able to use colors
\usepackage{amsmath} %Recomended for mathematical documents (use of: eqref,...)
\usepackage{amsfonts}
\usepackage{graphicx}
\usepackage{slashed}
\usepackage[dvipsnames]{xcolor}
\definecolor{refs}{RGB}{245,156,74}
\usepackage[colorlinks=true,hyperfootnotes=true,citecolor=cyan]{hyperref}

\usepackage{subcaption}
\usepackage{caption}
\usepackage{float}

\usepackage{physics}

\newcommand{\be}{\begin{equation}}
\newcommand{\ee}{\end{equation}}

\definecolor{tabblue}{HTML}{1f77b4}
\definecolor{taborange}{HTML}{ff7f0e}
\definecolor{tabgreen}{HTML}{2ca02c}
\definecolor{tabred}{HTML}{d62728}
\definecolor{tabpurple}{HTML}{9467bd}

\begin{document}

\title{Tail effects of self-interacting scalar fields}

\author{Philippe Brax}
\affiliation{Institut de Physique Th\'eorique, Universit\'e Paris-Saclay,CEA, CNRS, F-91191 Gif-sur-Yvette Cedex, France.}
\author{Emma Bruy\`ere}
\affiliation{Institut de Physique Th\'eorique, Universit\'e Paris-Saclay,CEA, CNRS, F-91191 Gif-sur-Yvette Cedex, France.}
\affiliation{Institut d'Astrophysique de Paris, UMR-7095 du CNRS et de Sorbonne Universit\'e, Paris, France.}

\begin{abstract}
    We consider the effects of quartic self-interactions on the dynamics of a binary system due to  a (nearly) massless scalar field conformally coupled to matter. We investigate the deviations from General Relativity at the conservative level and put a bound on the self-coupling $  \lambda \lesssim  (\beta^2 G_N M_\odot^2)^{-1}$ where $\beta$ is the conformal coupling of the scalar to matter. We also consider the radiative sector where we  use the Schwinger-Keldysh formalism to  find the tail interactions which couple the multipoles of the binary system and induce a small advance of the periastron. 
\end{abstract}

\maketitle

\section{Introduction}
	The lack of compelling explanation for the existence of both dark matter and dark energy, together with the possible gradual demise of the most popular models, i.e. the WIMPS (Weakly Interacting Massive Particles) for dark matter \cite{Arcadi:2024ukq} and the cosmological constant for dark energy \cite{Weinberg:1988cp,Copeland:2006wr}, might drive one to consider more seriously alternatives involving scalar fields. For dark matter, self-interacting light scalars may alleviate some of the small scale issues of the $\Lambda$-CDM model \cite{Hui:2021tkt}. For dark energy, the recent indication by the DESI collaboration in favour of dynamical dark energy might be the first sign that rolling scalars are on the move \cite{DESI:2024mwx}. In both cases, unless justified by a yet to be discovered new symmetry, these scalar should couple to matter. Such a coupling would have to be small as dark matter observations, e.g. the Bullet cluster, limit the possible self-interaction cross section \cite{Desjacques:2017fmf}. For dark energy, long range forces on cosmological scales would violate known properties of gravitation in the solar system~\cite {Bertotti:2003rm}. 

 In this paper, we will consider what is possibly the simplest scalar field system~\footnote{Of course, a simpler example is the case of a massless scalar with no self-interactions and a coupling to matter \cite{Damour:1992we}. The study of this case is well developed, see for instance \cite{Lang:2013fna,Lang:2014osa,Bernard:2018hta,Bernard:2018ivi,Bernard:2019yfz,Bernard:2023eul, Trestini:2024mfs}. The case of a massive scalar field is also well studied, see \cite{Huang:2018pbu, Diedrichs:2023foj}}. In this paper we work at the lowest order in the self-interaction coupling. As such  we focus on the example of a self-interacting scalar field, i.e. a massless scalar with quartic self-interactions\footnote{The cubic case at the conservative level is treated in \cite{Porto:2016pyg}. The existence of tail terms is also mentioned without any direct evaluation.}, and its Yukawa coupling to matter \cite{Damour:1992we}. We study the consequences of the self-interactions in both the conservative and radiative sectors of a binary system \cite{Goldberger:2004jt,Galley:2009px,Kuntz:2019zef,Brax:2019tcy,Goldberger:2022rqf}. We analyse these properties in the Schwinger-Keldysh formalism as this captures in one sweep and in a Lagrangian framework  the conservative and dissipative nature of the dynamics \cite{Galley:2009px,Galley:2012hx,Galley:2012qs,Galley:2015kus,Jakobsen:2022psy}. In the spirit of effective field theories \cite{Porto:2016pyg}, the quartic terms could be the most relevant effective scalar self-interactions on the scale of the binary system. They could emerge from the existence of a screening mechanism which reduces the coupling of the scalar to matter locally and involves non-linearity which could be  dominated  by the quartic term \cite{Brax:2021wcv}, or this could be the dominant term in the description of self-interacting dark matter with a range exceeding the size of the solar system \cite{Brax:2019fzb}. 

 At the conservative level, the self-interactions lead to a correction to Newton's law with a logarithmic running. This allows us to put bounds on the self-coupling $\lambda$ using the advance of perihelion of Mercury and the Shapiro time delay as constrained by the Cassini probe \cite{Will:2014kxa, Bertotti:2003rm}. Such a bound depends on the conformal coupling to matter $\beta$ and is tight in particular for dark matter models with very low masses~\cite{Brax:2019fzb}. In the radiative sector and after giving a new derivation in the Schwinger-Keldysh setting \cite{Schwinger:1960qe,Keldysh:1964ud,Chen:2017ryl} of the power emitted by scalars \cite{Galley:2009px,Kuntz:2019zef,Porto:2024cwd}, we focus on tail effects \cite{Blanchet:1987wq,Blanchet:1993ec,Blanchet:1994ez,Galley:2015kus,Porto:2024cwd,Foffa:2011np,Bernard:2016wrg,Foffa:2019eeb,Almeida:2021xwn,Trestini:2023wwg} as they naturally arise from the quartic self-interactions between the radiative and conservative sectors. The end result involves all the multipoles of the binary system which interact via an acausal kernel similar to the one in GR (General Relativity) \cite{Blanchet:1993ec,Galley:2015kus}.  Moreover they involve multiple time derivatives of the multipoles. We find that the scalar tail effects do not lead to a secular change of the orbits themselves as the averaged power in the tail interactions vanishes. On the other hand,  a small periastron advance is found \cite{Bernard:2016wrg}. They could only be of any significance if the solar system constraints could be relaxed in theories where the binary systems would couple strongly to the scalar field \cite{Ramazanoglu:2016kul}.  We leave the possible phenomenology of this effect to future work. 

 This paper is arranged as follows. In the first section \ref{point} we present the interaction between the scalar and the binary system in the Schwinger-Keldysh formalism. In section \ref{conservative} we consider the effects of the self-interactions on the conservative dynamics. In section \ref{radiative} we analyse the radiation sector and focus on the emitted power. In section \ref{tail} we calculate the tail effects for the self-interacting scalars. Finally we have added several appendices where technical details are given. 
	
 \section{Point-particles  in the Schwinger-Keldysh formalism} \label{point}

 \subsection{The conservative action }
	
We are interested in the dynamics of compact objects subject to both gravitational and scalar interactions. The latter is due to a light scalar field behaving like a nearly massless field in the environment of the objects under study. The dynamics of the system are influenced by the presence of the scalar field and its interaction with the compact objects. In a nutshell, the massive bodies generate a scalar field configuration which induces new forces between the objects. This back-reacts on the bodies' trajectories. In this work, we will concentrate on the conformal coupling of scalars to matter. We do not include a disformal part which would require a special treatment. 

We work in natural units $c=\hbar=1$ and we will use the metric signature $(-,+,+,+)$. For a single scalar field, the action reads 
\begin{equation}
    S[\varphi,J]=\int d^4x \left( -\frac{1}{2}\partial^{\mu}\varphi(x) \partial_{\mu}\varphi(x) - \frac{m^2}{2} \varphi^2 -V(\varphi) - J(x)\varphi(x) \right).
\end{equation}
 We will focus in this paper on the case $V(\varphi)=\lambda \varphi^4$. We will sometime introduce a mass term  with mass $m$ for the scalar field as an Infra-Red (IR) regulator and obtain that physical observables are finite in the small mass limit. As we are interested in the classical dynamics of compact objects interacting with gravitation and the scalar field, we will use as a first approximation the point-like limit where we take all the objects to be of vanishing size. This will introduce some UV divergences which will be regulated either with a momentum cut-off $\Lambda$ corresponding to the inverse of the largest size of the considered bodies $\Lambda=1/R$ or dimensional regularisation where the sliding scale $\mu$ can be also thought of as given by the typical size $1/R$, and should be determined by comparison with experiments. The associated renormalisation procedure will be discussed. 

 The point-like limit fits the effective field theory scheme used in this paper. Indeed, taken from afar, the objects appear point-like and the dynamics of the scalar field in this environment are taken to be the one of a nearly massless field with self-interactions. This could for instance result from the cosmological existence of light scalars acting as dark matter or dark energy whose local dynamics close to compact bodies reduce to the lowest order effective action, i.e. a quartic self-interaction complemented with a linear Yukawa coupling to matter. On larger distances as befitting the radiative part of the theory, the set of point-like sources will be replaced by a set of multipoles located at the centre of mass of the system. 
 We consider time-dependent point-like sources for the two body system located at $\vec x_{a,b}(t)$ which behave at the lowest order like
\begin{equation}
    J(y)=\frac{1}{M_{\varphi}}\big(m_a\delta^3(\vec{y}-\vec{x}_a(t_y))+m_b\delta^3(\vec{y}-\vec{x}_b(t_y))\big).
    \label{JJ}
\end{equation}
This expression will receive corrections from the velocity and  the Newtonian potential of the objects in the binary system which can be obtained systematically using the expansion scheme detailed below. 
Notice that $M_{\varphi}$ is a suppression scale related to the strength $\beta$ of the coupling of the scalar field to matter $\frac{1}{M_{\varphi}}=\frac{\beta}{m_{\rm Pl}}$  and $G_N=\frac{1}{8\pi m_{\rm Pl}^2}$ is Newton's constant.

The total action of the system can be derived in the Newtonian limit of General Relativity (GR) where the Einstein metric reads
\begin{equation}
    ds^2=-(1+2\Phi_N)dt^2 + (1-2\Phi_N)dx^2,
\end{equation}
and the particles representing the compact bodies follow the geodesics of the Jordan frame
\begin{equation}
    ds_J^2= e^{2\beta\varphi/m_{pl}}ds^2
\end{equation}  
where we denote by $g_{\mu\nu}$ the Einstein metric.
The total action of the system comprises the action due to the gravitational, scalar and matter fields with
	\begin{equation}
	S= \int d^4x \sqrt{-g}\left (\frac{R}{16\pi G_N}+ \frac{1}{2}\varphi\partial_{\mu}\partial^{\mu}\varphi - \frac{m^2}{2} \varphi^2 -\lambda\varphi^4  \right)  -m_a\int ds_a - m_b\int ds_b
	\end{equation}
 where we have introduced a mass $m$ to the scalar and a self-coupling $\lambda$. Notice that the gravitational and scalar terms are evaluated in the Einstein frame. The point particle actions are calculated with  the Jordan metric. The effective source term $J$ is obtained by expanding the matter actions and extracting the dependence on the scalar field. This is what we do below where we 
work at leading order in $G_N$ and in the non relativistic limit, $|\vec{v}_{a,b}|\ll 1$. The matter action is obtained at first order thanks to the infinitesimal interval in Jordan's frame
	\begin{equation}
	ds_J^2\simeq -(1+2\Phi_N+2\frac{\beta\varphi}{m_{pl}})dt^2 + (1-2\Phi_N+2\frac{\beta\varphi}{m_{pl}})dx^2
	\end{equation}
 corresponding to the Jordan frame action for the point particles
\begin{equation}
	S_a=-m_a\int ds_a \simeq -m_a \int dt \left( 1+ \Phi_{N,b}+ \frac{\beta\varphi_b(x_a)}{m_{pl}} - \frac{1}{2}v_a^2 \right).
\end{equation}
Here we have removed the matter self-interactions  which need to be renormalised and do not contribute to the dynamics of the binary system. Typically they are calculated using the dimensional regularisation scheme and therefore vanish. As a result we find that the scalar field couples to the current $J$ in (\ref{JJ}) and the particles are described by the Lagrangian of classical mechanics with a potential term coming from Newton's potential.

The effect of the scalar self-interactions in $\varphi^4$ can be taken into account by solving the Klein-Gordon equation perturbatively in $\lambda$
\begin{equation}
    \Box\varphi=J+ m^{{2}} \varphi+ 4\lambda\varphi^3 .
\end{equation} 
This allows one to obtain the scalar action
\be 
S_{\rm scalar}=\int d^4 x ( -\frac{1}{2} J \varphi + \lambda \varphi^4)
\ee
where the Klein-Gordon has been used. Here $\varphi$ satisfies the Klein-Gordon equation. 
The interaction potential depends on the solutions of the Klein-Gordon equation. This is given at leading order in $\lambda$ by $\varphi(x)=\varphi^{\lambda =0}_{J_{pp}}(x) $
where 
\begin{equation}
    \varphi^{\lambda =0}_{J_{pp}}(x)=i\int d^4y\Delta (x-y)J(y) \hspace{0,5cm}\text{and}\hspace{0,5cm} \Delta (x-y)= \int \frac{d^4p}{(2\pi)^4} \frac{ie^{-ip(x-y)}}{p^2+i\epsilon}.
\end{equation}
The propagator $\Delta(x-y)$ can be expressed with the Green function as $\Delta(x-y)=-iG(x-y)$, where the Green's function is a solution of $\square G = \delta(x-y)$. Here we have chosen the Feynman propagator which will be justified using the Schwinger-Keldysh formalism, see below. Moreover we have taken the mass $m$ to vanish as we will use it as an Infra-Red (IR) regulator taken to vanish in the end. Physically, we focus on systems much smaller that the Compton wave-length $1/m$ of the scalar. This is the typical situations for scalar dark matter and dark energy models as we will see below.

The solution to the Klein-Gordon equation can be expanded in powers of $\lambda$
\be 
\varphi(x)= \varphi^0(x)+ \sum_{n>0}^\infty \varphi^{(n)}
\ee
where $\varphi^0\equiv \varphi^{\lambda =0}_{J_{pp}}(x)$ and $\varphi^{(n)}(x)$ is of order $\lambda^n$. The first correction reads
\be 
\varphi^{(1)}(x)= 4\lambda \int d^4y G(x-y) (\varphi^{0}(y))^3.
\ee
In the following we consider the sources to be slowly moving, $|v_{a,b}|=|\dot{x}_{a,b}|\ll 1$ and since $p_{0}\sim m_{a,b}v_{a,b}^2$ and $|\vec{p}|\sim mv_{a,b}$, $p_{0}/\vert\vec{p}\vert \sim v_{a,b}\ll 1$, we can expand the propagator in powers of $p_{0}/|\vec{p}|$,
$
    \frac{1}{\vec{p}^2-p_0^2}=\frac{1}{\vec{p}^2} \left( 1 + \frac{p_0^2}{\vec{p}^2} + ... \right)
$.  
We get, as the leading term,
\begin{equation}
    \begin{aligned}
    \varphi^{0}(t,x) & = -\int d^4y\int \slashed{d}^4p\frac{e^{-ip(x-y)}}{p^2+i\epsilon} \frac{1}{M_{\phi}}\big(m_a\delta^3(\vec{y}-\vec{x_a}(t_y))+m_b\delta^3(\vec{y}-\vec{x_b}(t_y))\big) \\
    & = -\frac{1}{4\pi M_{\varphi}}\left( \frac{m_a}{|x-x_a(t)|} + \frac{m_b}{|x-x_b(t)|} \right) ,
    \end{aligned}
    \end{equation}
as already used and expected.  We will simplify the notation and denote this solution as $\varphi^0(t,x)=\varphi_a(t,x)+\varphi_b(t,x)$. 
The total action using the Poisson equation and the non-relativistic limit of the Einstein-Hilbert action becomes
\be 
S= \int dt (m_a \frac{v_a^2}{2}+m_b\frac{v_b^2}{2}-\frac{1}{2}(m_a \Phi_{N,b}+ m_b\Phi_{N,a})-\frac{1}{2}\frac{\beta}{m_{pl}}({m_b}\varphi_a(x_b)+{m_a}\varphi_b(x_a)) -V(x_a,x_b)).
\ee
The effective dynamics of the binary system are obtained by integrating out the Newtonian potential and the scalar field using their equations of motion. This is done by 
using  $\Phi_{N,b}=-\frac{G_Nm_b}{r}$, where $r=|\vec{r}|$ is the distance between the objects and  $\varphi_b(x_a)=-\frac{\beta m_b}{4\pi rm_{\rm Pl}}$ 
at leading order in $\lambda$.  This leads to  the action for  the particles in the binary system in the absence of self interactions 
\begin{equation}
	S_{\lambda=0}\simeq  \int dt \left( m_a \frac{v_a^2}{2}+m_b\frac{v_b^2}{2}-\frac{1+2\beta^2}{2}(m_a \Phi_{N,b}+ m_b\Phi_{N,a})  \right).
\end{equation}
where Newton's constant is multiplied by $1+2 \beta^2$ due to the exchange of the massless scalar between the two point particles.

The induced interaction potential due to the self-interactions is given by\footnote{
The contribution from $\varphi^{(1)}(x)$ in the action follows from \be 
-\int d^4x \frac{1}{2}J(x)\varphi^{(1)}(x)= -2\lambda \int d^4x (\varphi^0(x))^4
\ee
where we have used that the Feynman propagator satisfies $G(x)=G(-x)$.
}
\be 
V(x_a,x_b)= \lambda \int d^3x  (\varphi^0 (\vec x,t))^4.
\ee
This potential describes the effect of the scalar self-interactions on the dynamics of the binary system, i.e. they induce a new potential term which modifies the trajectories of the compact objects. 
At leading order in $\lambda$, this provides the correction from the scalar self-interactions to the conservative part of the system. The radiative part associated to dissipation by scalar waves emitted to infinity will be described below. 
    
\subsection{The Schwinger-Keldysh description}
 
So far we have only described the leading effects of the scalar self-interactions on the dynamics of the binary system. As well-known, the binary system also radiates and the resulting scalar field carries energy to infinity. Taking into account the effect of radiation can be easily incorporated in the description of the system. A first approach consists in calculating the radiated flux 
\be 
P= \int d^2S n_i \partial^0 \varphi \partial^i\varphi
\label{flux}
\ee
on a large sphere at infinity \cite{Ross:2012fc}. This requires solving the Klein-Gordon far away from the binary system for scalar waves with $p_0\sim \vert \vec p\vert$ using the retarded Green's function. 
Another approach consists in calculating the scalar action in the "in-out" formalism where the propagators are of the Feynman type and the imaginary part of the action is picked. This corresponds to the exponential decay of the evolution operator of the system from which the dissipative part can be extracted \cite{Kuntz:2019zef}. 

The main difficulty of treating dissipation in a Lagrangian formulation is the non-conservation of energy for the emitting system, i.e. the binary system. This can be remedied using  
the Schwinger-Keldysh ("in-in") formalism where a "shadow" scalar field is introduced whose physical purpose is to re-inject energy from infinity back into the system. With this, a Lagrangian formulation of dissipation can be treated \cite{Galley:2012hx}. 

\begin{figure}[ht] 
	\includegraphics[scale=0.5]{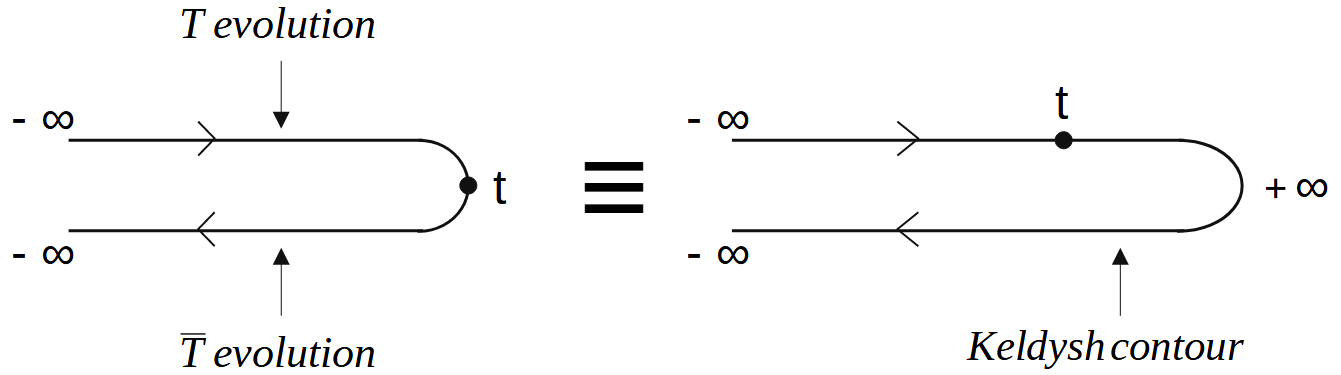}				
	\caption{Keldysh contour with an operator inserted at time $t$ and the contour rejected to infinity thanks to  $U^\dag(t,+\infty)U(t,+\infty)=I$.}
  \label{k}
	\end{figure}

Let us consider the scalar field theory with its self-interactions as a quantum system. Contrary to the "in-out" formalism, we focus on vacuum expectation values of the scalar field and its time evolution, see \cite{Donath:2024utn} for a modern take on the similarities and differences between the two formulations. When coupled to the binary system, this allows one to take into account the radiation of scalars to infinity.
In the interaction picture, where the interaction Hamiltonian is $H_I = \int d^3x V(\varphi_I)$, the evolution operator is $U=T\left( \exp(-i \int_{-\infty}^t dt' H_I(t')) \right)$. The full quantum field $\phi$ is related to the field in the interaction picture $\phi_I$ as $\phi=U^\dag \phi_I U$, with $U^\dag =\bar{T}\left( \exp(i \int_{-\infty}^t dt' H_I(t')) \right)$. Here $T$ is the T-product that orders the operators with increasing time from right to left and $\bar{T}$ orders the path with decreasing time from right to left too. Defining by $\ket{\rm in}$ the vacuum of the free field theory, the expectation values of the quantum field are such that
    \begin{equation}
    \langle {\rm in}\vert \phi\vert {\rm in}\rangle  =  \langle \rm{in}\vert \bar{T}\left( e^{i \int_{-\infty}^t dt' H_I(t')} \right) \phi_I T\left( e^{-i \int_{-\infty}^t dt' H_I(t')} \right)\vert \rm{in}\rangle     
    \end{equation}
Hence by  reading the evolution of the operators from right to left, we can draw the diagram \ref{k} where the operators in the first $T$ product are on the top line and the ones with the $\bar{T}$ product on the bottom line. More generally, we can introduce the generating function of correlators in the "in-in" formalism, $Z[J]=\langle {\rm in}\vert  T\left( e^{-i \int d^4x J\phi} \right)\vert {\rm in} \rangle$. This can be written as a path integral with  a field evolving with increasing time (to the right, $T$) and another with decreasing time (to the left, $\bar{T}$).	
We assign the subscript 1 to the quantities following the $T$ evolution on the top contour and 2 to the ones following the $\bar{T}$ evolution \cite{Schwinger:1960qe, Keldysh:1964ud}
    \begin{equation}    
    Z[J]=\int \mathcal{D}\varphi_1 \mathcal{D}\varphi_2 e^{iS_1-i\int d^4x J_1\varphi_1 - iS_2+i\int d^4x J_2\varphi_2}
    \end{equation}
    where $S_{1,2}=S(\varphi_{1,2})$.
    Classically the sources $\bar J_{1,2}$ are assigned to the top and bottom contours too.  We have also $\varphi_1(+\infty)=\varphi_2(+\infty)$ by continuity.  Now we shift $J \rightarrow \bar{J} + \delta J$, where $\bar{J}$ is the classical source that we considered in the previous sections and the variation $\delta J$ is associated to quantum processes.
    As in usual quantum field theory we introduce the generating function of connected diagrams $W$, $Z[J]=e^{iW[J]}$, so here,
    \begin{equation}
    Z[J_1,J_2]=e^{iW[J_1,J_2]}
    \end{equation}
    and we perform a Legendre transform to obtain the generating function of one-particle irreducible diagrams
    \begin{equation}
    W[J_1,J_2] = \Gamma[\Phi_1,\Phi_2] - \int d^4xJ_1\Phi_1 +\int d^4x J_2\Phi_2,
    \end{equation}
    where the  fields are obtained as
    \begin{equation}
    \Phi_1=-\frac{\delta W}{\delta J_1} \hspace{0,5cm}\text{and}\hspace{0,5cm} \Phi_2=\frac{\delta W}{\delta J_2},
    \end{equation}
    such that the equations of motion for the $\Phi_{1,2}$ fields are
    \begin{equation}
    \frac{\delta\Gamma}{\delta\Phi_1}=J_1 \hspace{0,5cm}\text{and}\hspace{0,5cm} \frac{\delta\Gamma}{\delta\Phi_2}=-J_2 .
    \end{equation}
    When $\delta J_{1,2}=0 \Rightarrow \Phi_{1,2}=\bar{\Phi}_{1,2}$ which is the classical solution of the quantum field theory. At the classical level, this corresponds to the calculation we presented in the previous section with external sources involving the binary system. We denote these solutions by $\bar \Phi_{1,2}$.

\subsection{The classical dynamics}

    We can now formally obtain the classical action for the point-particles after integrating out the scalar field, i.e. solving its classical equations of motions. This gives a correction to the matter action and therefore corrects Newton's law for non-relativistic particles.
    The matter action in the Schwinger-Keldysh formalism is also obtained by doubling the number of fields and introducing the trajectories of shadow objects. We denote by $ x^{a,b}_{1,2}$ the positions of the objects and their shadows. The matter action is simply
    \be 
    S_{\rm matter}= \int dt_1 {\cal L}^0(x_1,\dot x_1)- \int dt_2 {\cal L}^0(x_2,\dot x_2)
    \ee
    where 
    \be 
    {\cal L}^0(x_{1,2},\dot x_{1,2})= \frac{1}{2}(m_a (v_{1,2}^a)^2+ m_b (v_{1,2}^b)^2) +\frac{G_Nm_am_a}{(x_{1,2}^a-x_{1,2}^b)^2}
    \ee
    and $v^{a,b}_{1,2}= \dot x^{a,b}_{1,2}$ is the velocity.
This classical action is corrected by the scalar action integrated over all the field configurations which reduces at the classical level to solving the Klein-Gordon equation with sources from the compact bodies. This can be written as
    \begin{equation}
    S_{\rm eff}[x_1,x_2]\equiv \Gamma[\bar\Phi_1,\bar \Phi_2] - \int d^4x\bar J_1\bar \Phi_1 +\int d^4x \bar J_2\bar \Phi_2=\int dt\left( \mathcal{L}[x_1]-\mathcal{L}[x_2]+R[x_1,x_2] \right)
    \end{equation}
    where $\mathcal{L}$ accounts for the binary system's  conservative
interactions while $R$ corresponds to the  non-conservative effects, such as radiation reaction.  In the low velocity approximation that we follow, the functions ${\cal L}$ and $R$ are only dependent on $x_{1,2}$. The fields $\bar\Phi_{1,2}$ comprise both the conservative and the radiative parts and lead to the conservative, dissipative and also tail effects where the conservative and radiative parts of the scalar field interact. 

The classical equations of motion for the compact bodies are easier to obtain by changing basis and defining
\be 
x_-= x_1-x_2, \ \ x_+= \frac{x_1+x_2}{2}
\ee
The physical equations are obtained by taking the variation of the action with respect to $x_-$ and setting  $x_-=0\Rightarrow x_+=x_1=x_2=x$ afterwards.
The matter action reduces to 
\be 
S_{\rm matter}= \int dt ( m_a v_-^a v_+^a + m_b v_-^b v_+^b +\frac{G_N m_a m_b}{\vert x_+^a-x_+^b +\frac{1}{2}(x_-^a-x_-^b)\vert }-\frac{G_N m_a m_b}{\vert x_+^a-x_+^b -\frac{1}{2}(x_-^a-x_-^b)\vert })
\ee
Taking the variation of $S_{\rm matter}+ S_{\rm eff}$ with respect to $x_-$  and putting $x_-=0$ we obtain Newton's 
\be 
m_{a,b}\ddot x^{a,b}= - \frac{\partial\Phi_N}{\partial x^{a,b}} +\frac{\delta S_{\rm eff}}{\delta x_-^{a,b}}\biggr\arrowvert_{x_-^{a,b}=0}
\ee
where Newton's potential is $\Phi_N=-\frac{G_N m_a m_b}{\vert x^a-x^b\vert }$.
The contribution from the scalar effective action reads
\be 
\frac{\delta S_{\rm eff}}{\delta x_-^{a,b}}\biggr\arrowvert_{x_-^{a,b}=0}=
\frac{\partial {\cal L}[x^{a,b}]}{\partial x^{a,b}} + \frac{1}{2}\left(
\frac{\partial R[x_1,x_2]}{\partial x_1}-\frac{\partial R[x_1,x_2]}{\partial x_2}\right)\biggr\arrowvert_{x_1=x_2}
\ee
The first term is simply a potential contribution to Newton's law and in the previous section we identified ${\cal L}\equiv -V$ in the case of a quartic interaction. We will confirm this below. The remaining interaction involving $R$ is the one due to non-conservative effects. We will see that the scalar dissipation and tail effects appear of this form. 

\section{The point-particle Wilsonian action} \label{conservative} 
\subsection{Integrating out short distances}    
We will obtain the point-particle action in the Schwinger-Keldysh formalism in a two step approach. First we will integrate out over the field configurations on short distances corresponding to the size of the binary system, see appendix \ref{app:eff} for details. This will give rise to a Wilsonian action for the long distance fields which describe the radiation from the binary system. In short, we decompose the field as
\be 
\varphi=\underbrace{\varphi_0}_{\text{short distances}} + \underbrace{\phi}_{\text{large distances}}
\ee
corresponding to  the conservative and the radiation parts associated to the short and  large distance fields. This is reminiscent of  the usual decomposition between high and low energy field in a Wilsonian treatment of quantum field theory where the short distance physics is eventually integrated out. Let us start with the generating functional 
\begin{equation}
	\begin{aligned}
	\int \mathcal{D}\varphi_1 \mathcal{D}\varphi_2 e^{iS_1(\varphi_1) - iS_2(\varphi_2) -i\int d^4 x  J_1\varphi_1 +i\int d^4x J_2\varphi_2} & 
	& = \int \mathcal{D}\phi_{1} \mathcal{D}\phi_{2} e^{iW(\phi_1,\phi_2,J_1,J_2) -i\int d^4 x  J_1\phi_1 +i\int d^4 x J_2\phi_2} 
	\end{aligned}
\end{equation}	
where $J=\bar J+\delta J$. The connected generating functional is given by 
\be 
e^{iW(\phi_1,\phi_2,J_1,J_2)} =\int \mathcal{D}\varphi_{1,0} \mathcal{D}\varphi_{2,0}  e^{iS_1(\varphi_{1,0}+\phi_1) - iS_2(\varphi_{2,0}+\phi_2) -i\int J_1\varphi_{1,0} + i\int J_2\varphi_{2,0}} 
\ee
We can introduce the effective action $\Gamma$ as the Legendre transform with respect to the conservative field
    \begin{equation}
    W(\phi_1,\phi_2,J_1,J_2) = \Gamma[\phi_1,\phi_2,\Phi_{1,0},\Phi_{2,0}] -\int d^4x  J_1\Phi_{1,0} +\int d^4x J_2\Phi_{2,0}
    \end{equation}
where $\Phi_{1,0}=-\frac{\delta W}{\delta J_1}, \Phi_{2,0}= \frac{\delta W}{\delta J_2} $.
At the classical level where  $\delta J_{1,2}=0$  the classical conservative fields are given by $\bar{\Phi}_{1,0}=-\frac{\delta W}{\delta J_1}\arrowvert_{\delta J_1=0}$ and $\bar{\Phi}_{2,0}=\frac{\delta W}{\delta J_2}\arrowvert_{\delta J_2=0}$. They also solve the classical field equations $\frac{\delta \Gamma}{\partial \Phi_1}\vert_{\bar \Phi_{1,2}}=\bar J_1 $ and  $\frac{\delta \Gamma}{\partial \Phi_2}\vert_{\bar\Phi_{1,2}}=-\bar J_2 $. It is important to notice that the classical solutions $\bar \Phi_{1,2}(\phi_{1,2})$ depend on the fields $\phi_{1,2}$. 

The effective action for the point-particles obtained by integrating over the scalar fields is given by
    \begin{equation}
    e^{iS_{\rm eff}[x_1,x_2]}=\int \mathcal{D}\phi_{1} \mathcal{D}\phi_{2} e^{i\Gamma(\phi_1,\phi_2,\bar{\Phi}_{1,0},\bar{\Phi}_{2,0}) -i\int \bar{J}_1\bar{\Phi}_{1,0} +i\int  \bar{J}_2\bar{\Phi}_{2,0}-i\int d^4 x  \bar J_1\phi_1 +i\int d^4 x \bar J_2\phi_2} .
    \end{equation}
    At the classical level, the one particle irreducible effective action reduces to the classical action evaluated around the classical solution of the equations of motion for the short distance fields, 
\be 
\Gamma(\phi_1,\phi_2,\bar{\Phi}_{1,0},\bar{\Phi}_{2,0}) = S_1(\phi_1+\bar{\Phi}_{1,0}) - S_2(\phi_2+\bar{\Phi}_{2,0}).
\ee
We can now Taylor expand the action and obtain 
    \begin{equation} \label{clfi}
    \begin{aligned}
    S_1(\phi_1+\bar{\Phi}_{1,0}(\phi_1)) = S_1(\bar{\Phi}_{1,0}(0)) + \int d^4 x \frac{\delta S_1(\phi_1+\bar{\Phi}_{1,0}(\phi_1))}{\delta\phi_{1}}\biggr\arrowvert_{\bar{\Phi}_{1,0}(0)}\phi_1 + \underbrace{\delta S_1(\phi_1,\bar{\Phi}_{1,0})}_{\text{higher order}} 
    \end{aligned}
    \end{equation}
The first term yields the conservative part of the effective action for the point-particles. The second term defines the effective source of the radiative field
\be 
{\cal J}^\lambda= \frac{\delta S(\phi+\bar\Phi(\phi))}{\delta \phi}\biggr\arrowvert_{\bar \Phi(0)}
\ee
which contains the effects of the conservative system on the radiative system at the linear level.
The last term $\delta S$  encodes the non-linear self-interactions of the radiative field and its interaction with the conservative part. For a scalar field with an interacting potential $V_{\rm int}(\phi)$, we have, see the appendix \ref{app:eff} for more detail
\be 
{\cal J^\lambda}= - \frac{\partial V_{\rm int}}{\partial \phi}\biggr\arrowvert_{\bar \Phi(0)}.
\ee
The total source term is therefore 
\be 
{\cal J}={\cal J}^\lambda +{\cal J}_{\rm matter}
\ee
where ${\cal J}_{\rm matter}= -\bar J.$

Viewed from far away, the binary system appears as a point-like source whose characteristics are summarised by the multipole expansion
\begin{equation}
    \int d^4x {\cal J}(x)\phi(x)= \int dt d^3x {\cal J} (\vec x,t)\sum_n \frac{1}{n!} x^{i_1}\cdots x^{i_n} \partial_{i_1}\cdots\partial_{i_n} \phi(\vec 0,t) = \sum_n \frac{1}{n!} \int dt {\cal J}^{i_1 \cdots i_n} \partial_{i_1}\cdots\partial_{i_n} \phi(\vec 0,t).
\end{equation}
where the centre of mass of the binary system has been put at the origin.
This can be encapsulated using the mixed Fourier representation
    \begin{equation}
    {\cal J}(t,\vec{k})=\int d^3\vec{x} e^{-i\vec{k}.\vec{x}}{\cal J}(t,\vec{x}) = \sum_n \frac{(-i)^n}{n!} J^{(n)}_{i_1 \cdots i_n}(t)k^{i_1}\cdots k^{i_n}.
    \end{equation}   
In the following we will rearrange the multipole expansion and will consider the trace free multipole $I^{i_1\dots i_n}$.
For the monopole, dipole and quadrupole they read
    
\begin{equation}
I= \int d^3 x ({\cal J} + \frac{1}{6}\ddot {\cal J} x^2),\ I^i= \int d^3 x x^i {\cal J}, \ I^{ij}=\int d^3x Q^{ij} {\cal J}
\end{equation}
where
\be 
Q^{ij}= x^i x^j - \frac{1}{3} x^2 \delta^{ij}.
\ee
Here $x^i$ denotes the coordinates around the centre of mass. We will compute these multipoles explicitly for the case of  quartic self-interactions and a linear coupling to matter

Finally the non-linear term in $\phi_{1,2}$ correspond to  the radiative part of the scalar action. 
The final integration over the radiative field is done at tree level, i.e. classically, by solving the two Klein-Gordon equation for $\phi_{1,2}$ sourced by the multipoles of the binary system. This will be done in detail below. 
    
\subsection{Conservative part}    
    
We can now carry out the integration over the short distance field explicitly. This will lead to the conservative part of the point-particle effective action. The Klein-Gordon equation for the classical fields can be solved by iteration, here for $\bar \Phi_1$, 
    \begin{equation}
    \begin{aligned}
    \left\lbrace \begin{array}{l}
	\square \bar{\Phi}_{1,0}^{} =  \bar{J}_1+ m^2 \bar{\Phi}_{1,0}^{} + 4\lambda\left(\bar{\Phi}_{1,0}^{}\right)^3 
\end{array}\right.
    \end{aligned}
    \end{equation}
where the complete solution is given in perturbation in $\lambda$ by  $\bar{\Phi}_{1,0} = \sum_{n=0}^{\infty} \bar{\Phi}_{1,0}^{(n)}$. At each step the Green's functions with the appropriate boundary conditions for the two fields $\bar\Phi_{1,2}$ must be used.  In the Schwinger-Keldysh formalism, the Green's functions satisfy
\begin{equation}
    (\square - m^2) G_{ab} = \delta(x-y)c_{ab}
\end{equation}
with $c_{11}=1$, $c_{22}=-1$ and $c_{12}=c_{21}=0$.
Here $G_{11}=G_F$ is the Feynman Green's function, $G_{22}=G_F^*$,  $G_{12}$  is the negative frequency part whilst $G_{21}$ is the positive frequency part. Explicitly and in Fourier space we have 
\begin{equation}
    \left\lbrace \begin{array}{ll}
	G_{11}= \frac{1}{\omega^2-\vec p^2-m^2+i\epsilon} & G_{12}= \theta(-\omega)\delta(\omega^2-\vec p^2-m^2) \\
	G_{21}= \theta(\omega)\delta(\omega^2-\vec p^2-m^2) & G_{22}= \frac{1}{\omega^2-\vec p^2-m^2-i\epsilon}
\end{array}\right.
\end{equation}
So for the conservative  field we obtain
\begin{equation}
    \bar{\Phi}_{1,0}^{(0)}(x) = \int d^4x' \left( G_{11}(x-x')\bar{J}_1(x') + G_{12}(x-x')\bar{J}_2(x') \right).
\end{equation}
The second term vanishes since we have in Fourier space, $\theta(-\omega)\delta(\omega^2-\vec p^2-m^2)\delta(\omega)J(\vec p)=0$.\
Thus we obtain the same expression as used previously for the $0^{th}$ order solution of the Klein-Gordon equation using the Feynman propagator. We can now evaluate
\begin{equation}
 {\cal L}[x^{a,b}]=  \int d^3x \left( \frac{1}{2}\bar{\Phi}\square\bar{\Phi} - \frac{1}{2}m^2\bar{\Phi}^2 -\lambda\bar{\Phi}^4 - \bar{J}\bar{\Phi}\right) =\int d^3x \left( \lambda\bar{\Phi}^4 - \frac{1}{2} \bar{J}\bar{\Phi}\right). 
\end{equation}
At leading order this reduces to
\begin{equation}
 {\cal L}[x^{a,b}]=  \int d^3x \left( -\lambda({\bar{\Phi}^{(0)}})^4 - \frac{1}{2} \bar{J}\bar{\Phi}^{(0)}\right). 
\end{equation}
The matter part depending on $\bar J$ is the one responsible for the correction $G_N\to G_N(1+2\beta^2)$ in Newton's potential, i.e. the scalar interaction increases the strength of the gravitational interaction. The quartic term is the potential due to the scalar self-interactions that we will calculate below. 

\subsection{The self-interaction potential}
The quartic self interaction in the scalar action at first order in $\lambda$ provide us with the interaction potential for the point-particles. After using the equations of motion we have for the potential term
\begin{equation}
	\int d^4x \lambda(\bar \Phi^{(0)})^4 = \frac{\lambda}{(4\pi M_{\Phi})^4} \int dt d^3x \left( {6}\frac{m_a^2m_b^2}{|x-x_a|^2|x-x_b|^2} +4\frac{m_a^3m_b}{|x-x_a|^3|x-x_b|} + 4\frac{m_am_b^3}{|x-x_a||x-x_b|^3} \right).
\end{equation}
 We only take into account the terms mixing the two sources as they are the only ones influencing the dynamics of the binary system.  The prefactor is also given by  $\frac{1}{(4\pi M_{\Phi})^4}=\frac{\beta^4G_N^2}{4\pi^2}$. Using that $\frac{1}{|x|^2}=2\pi^2\int \slashed{d}^3 p \frac{e^{i\vec{p}.\vec{x}}}{|\vec{p}|}$, we obtain
	\begin{equation}
	\int d^3x \frac{1}{|x-x_a|^2|x-x_b|^2} = \int d^3x (2\pi^2)^2 \int \slashed{d}^3p \slashed{d}^3k \frac{e^{i(\vec{p}+\vec{k}).\vec{x}-i\vec{p}.\vec{x}_a-i\vec{k}.\vec{x}_b}}{|\vec{p}||\vec{k}|} = 4\pi^4  \int {\slashed{d}^3p } \frac{e^{-i\vec{p}.\vec{r}}}{|\vec{p}|^2} = \frac{\pi^3}{r} .
	\end{equation}
Let us now compute the  other terms. We define $f(p)$ such that $\frac{1}{|x|^3}=\int \slashed{d}^3p  f(p)e^{i\vec{p}.\vec{x}}$ then
\begin{equation} \label{cut}
	\begin{aligned}
	f(p) = \int d^3x \frac{e^{-i\vec{p}.\vec{x}}}{|x|^3}=\lim_{\epsilon\rightarrow 0} \frac{2i\pi}{p}\int_{\epsilon}^{\infty}dx \frac{e^{-ipx}-e^{ipx}}{x^2}
	\end{aligned}
\end{equation}
where we  introduce and arbitrary  cutoff $\epsilon=1/\Lambda $ due to a UV divergence, i.e. we put a limit on small distances at $\epsilon$. 
Then we obtain
\begin{equation}
	\begin{aligned}
	\int_{\epsilon}^{\infty}dx \frac{e^{-ipx}-e^{ipx}}{x^2}=-ip\left( \Gamma(0,-i\epsilon p) + \Gamma(0,i\epsilon p) \right) - {2}i \frac{\sin(\epsilon p)}{\epsilon} 
	\end{aligned}
\end{equation}
	where $\Gamma(s,z)=\int_z^{\infty}t^{s-1}e^{-t}dt$ is the incomplete gamma function and $\Gamma(0,z)=-\gamma-\ln(z)-\sum_{k=1}^{\infty} \frac{(-z)^k}{k(k!)}$. So for small $\epsilon$, we have 
$
	f(p) = -2\pi\left( 2\gamma - {2} + \ln(\epsilon^2 p^2) \right).
$
We can now compute the other terms using this result
	\begin{equation}
	\int d^3x  \frac{1}{|x-x_a||x-x_b|^3}= 2\pi\frac{({2}-2\gamma)}{r} - 4 \int_0^{\infty} \frac{du}{r} \frac{\sin(u)}{u}\ln(\frac{\epsilon^2 u^2}{r^2}) = \frac{{4}\pi+{2\pi}\ln(r^2/\epsilon^2)}{r}
	\end{equation}
where we used that $\int_0^{\infty} du \frac{\sin(u)}{u}\ln(u^2)=-\gamma\pi$.
Thus we obtain for the potential due to the self-interactions
from $\int d^4x \lambda(\bar \Phi^{(0)})^4$ as 
\begin{equation}
V(r)=	 \frac{\lambda\beta^4G_N^2}{4\pi^2} \left( \frac{{6}\pi^3m_a^2m_b^2}{r} + 4m_am_b(m_a^2+m_b^2)\frac{({4}\pi+{2\pi}\ln(r^2\Lambda^2))}{r}  \right) .
\end{equation}
This is a repulsive correction to Newton's law with a logarithmic running coming from the UV divergence due to the point-like nature of the interacting objects. As the physical observables cannot depend on the arbitrary cut-off scale $\Lambda$, the result must be renormalised.

\subsection{Renormalisation}
Although we are dealing with a classical theory, we have just seen that the potential generated by the scalar self-interactions is sensitive to the UV cut-off, i.e. the short distance physics. This cut-off scale is arbitrary and corresponds to considering that the description of the binary system is valid only on distance far enough $r\gtrsim 1/\Lambda$. Eventually, this long distance behaviour should be matched with the UV description of the binary system, i.e. the actual dependence of the classical trajectories on the details of the distribution of matter inside each object. The long distance description using point-like particles and the short distance one depending on the finite size effects should match at a scale $R_M=1/\Lambda_M$. The matching at $R_M$ serves as renormalisation condition for the long distance physics as we will see below.  

The cut-off dependence of the UV divergence can be removed by adding a counter-term to the action of the form \cite{Mougiakakos:2020laz}
\be 
\delta S= 2 c(\Lambda)\int d^4x \sqrt {-g} \frac{R}{16\pi G_N} \frac{\phi}{m_{\rm Pl}}
\ee
where this non-minimal operator depends on the coupling $c(\Lambda)$ which is a function of  the cut-off. Using the Einstein equations, the Ricci scalar can be identified with $R=-8\pi G_N T$ where $T$ is the trace of the energy momentum tensor of matter, i.e. in the non-relativistic limit
\be 
T= -m_a \delta^{(3)}(x-x_a) - m_b \delta^{(3)}(x-x_b).
\ee
This implies that the on-shell contribution of this non-minimal operator to the potential is 
\be 
\delta V=- \frac{c(\Lambda)}{m_{\rm Pl}}(m_a \phi^0(x_a)+ m_b\phi^0(x_b))
\ee
where the coinciding point divergences are discarded as they give irrelevant constants which do not influence the equations of motion. This corresponds to
\be 
\delta V(r)= \frac{4 G_N m_a m_b \beta c(\Lambda)}{r}
\ee 
implying that
\be 
c(\Lambda)= -\frac{\lambda \beta^3 G_N (m_a^2 +m_b^2)}{\pi}\ln \Lambda
\ee
up to a constant which will be fixed by the renormalisation condition. The original action is written in the Einstein frame with a coupling to matter associated to the Jordan metric
$g^J_{\mu\nu}= A^2(\phi) g_{\mu\nu}$ where $A^2(\phi)\sim 1+ 2 \beta \frac{\phi}{m_{\rm Pl}}$. The non-minimimal coupling corresponds to a change of Jordan frame with a new scale-dependent coupling and the new Einstein frame is such that $g_{\mu\nu}= B^2(\phi) g_{\mu\nu}^E$ where $B^2(\phi) \sim 1-2c(\Lambda) \frac{\phi}{m_{\rm Pl}}$ implying that the total change of frame between the Jordan to the Einstein frame is determined by 
\be 
\beta(\Lambda)= \beta - c(\Lambda)
\ee
where the large logarithms can be summed using the renormalisation group equation
\be 
\frac{d\beta}{d\ln \Lambda}= -b_3 \beta^3
\ee
where $b_3= -\frac{\lambda G_N (m_a^2 +m_b^2)}{\pi}$. The solution is simply
\be 
\beta^2(\Lambda)= \frac{\beta^2_M}{1+ b_3 \beta_M^2 \ln \frac{\Lambda^2}{\Lambda_M^2}}
\ee
where we have used the matching, i.e. the renormalisation, prescription
\be 
\beta(\Lambda_M)\equiv \beta_M.
\ee
This corresponds to the perturbative expansion
\be 
\beta(\Lambda)= \beta_M - b_3 \beta_M^3 \ln \frac{\Lambda}{\Lambda_M}.
\ee
The running coupling $\beta(\Lambda)$ corresponds to the effective coupling of the scalar field to matter when the effects of the short distances $r\lesssim 1/\Lambda$ have been integrated out. The running coupling would have a Landau pole for large  values of $\Lambda$  although this is beyond the matching scale $\Lambda_M$ where the appropriate description of the binary system, i.e. its UV completion, with finite size objects should be used. Similarly in the IR regime when $\Lambda \to 0$ the effective coupling decreases, i.e. a classical analogue of asymptotic freedom.  We will see that the classical tests of GR imply that $b_3\beta_M^2$ is small for binary systems of solar masses implying that the renormalisation effects are small in this case.  
Notice that after renormalisation the potential becomes
\begin{equation}
V(r)=	 \frac{\lambda\beta_M^4G_N^2}{4\pi^2} \left( \frac{{6}\pi^3m_a^2m_b^2}{r} + 4m_am_b(m_a^2+m_b^2)\frac{({4}\pi+{2\pi}\ln(r^2\Lambda_M^2))}{r}  \right) .
\end{equation}
where the logarithmic corrections involves the matching scale $\Lambda_M$. 
\subsection{The effective one-body metric and GR tests}
In the following, we denote by $\Lambda$ the matching scale $\Lambda_M $ and $\beta$ the renormalised coupling $\beta_M$ for convenience sake. {We extract the effective metric and later study GR tests \cite{Brax:2018bow}.}
In the centre of mass frame, the Lagrangian associated to the motion of the reduced particle of mass $\mu= \frac{m_am_b}{m_a+m_b}
$ in the gravitational field of the total mass $M= m_a+m_b$	is given by
\begin{equation}
	\frac{\mathcal{L}}{\mu} = \frac{1}{2} v^2 +\frac{G_N M(1+2\beta^2)}{r} - \frac{\lambda\beta^4G_N^2M^3}{r} \left( {\frac{3}{2}}\pi\nu  +\frac{{4}(1-2\nu)}{\pi} + \frac{{2}(1-2\nu)}{\pi} \ln \Lambda^2 r^2\right) 
\end{equation}	
where $\nu=\mu/M$. At leading order in $\lambda$, this lagrangian can be reformulated using an effective one-body metric $g^{\rm eff}_{\mu\nu}$ corresponding to the geodesic motion of a particle of mass $\mu$ affected by both gravitation and the scalar self-interaction \cite{Julie:2017pkb}, i.e. 
\begin{equation}
	\mathcal{L}^{\text{eff}}= -\mu\sqrt{-g_{\mu,\nu}^{\text{eff}}u^{\mu}u^{\nu}},
\end{equation}
where $u^\mu=\frac{dx^\mu}{d\tau}$ and $\tau$ is the proper time of the particle along its motion.
The components of the metric correspond to perturbations of the initial Jordan metric
\begin{equation}
	\begin{aligned}
	& g^{\text{eff}}_{00} = -\left[ 1 + 2\Phi_N\left( 1 + 2\beta^2 - \lambda\beta^4G_NM^2 A \right) + \lambda\beta^4G_N^2 M^3 B\frac{\ln(\Lambda^2r^2)}{r} \right] \\
	& g^{\text{eff}}_{ij} = \left[ 1 - 2\Phi_N(1-2\beta^2)\right] \delta_{ij}
	\end{aligned}
\end{equation}
defining $A= \left( {\frac{3}{2}}\pi\nu +\frac{{4}(1-2\nu)}{\pi} \right)$ and $B= \frac{4(1-2\nu)}{\pi}$ for convenience and the Newtonian potential is due to the central mass being $M$, i.e. $\Phi_N=-\frac{G_NM}{r}$. 
The dynamics of the binary system are now  reduced to the one of an object of mass $\mu$  and its interaction with the scalar field is now encoded in a perturbation to the spatial part of the metric. 
	
\subsubsection{The perihelion advance}
	
The perihelion advance of a small object such as Mercury orbiting around a larger one such as the Sun is a classic test of GR. It corresponds to the angular shift of the perihelion position due to perturbations to the interaction between the bodies that are not taken into account in a classical Newtonian two body system.	
	\begin{figure}[ht]
	\includegraphics[scale=0.5]{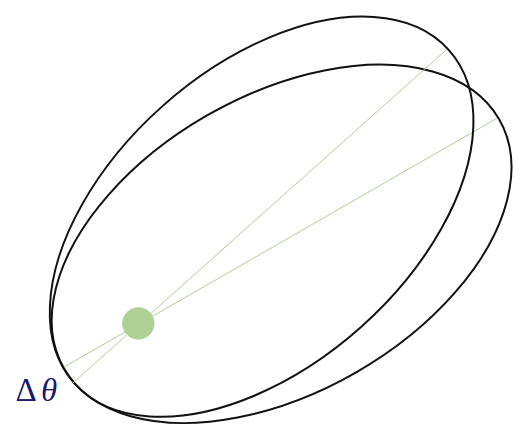}				
	\caption{Representation of the advance of the perihelion corresponding  the angular shift of the perihelion.}
	\end{figure}
We parameterise the trajectories with the proper time $\tau$ such that $
g_{\mu,\nu}^{\text{eff}}u^{\mu}u^{\nu} =-1$. 
We analyse the motion  with polar coordinates in the orbital plane, $d\Omega^2\equiv d\theta^2$. Since there is no explicit dependence on $\theta$ and $t$ in the effective Lagrangian there are two  conserved quantities from the Euler-Lagrange equations
	$J = \frac{\partial\mathcal{L}^{\text{eff}}}{\partial\dot{\theta}} = \mu g_{\theta\theta}^{\text{eff}} \dot{\theta}$ and $	 -\mu k = \frac{\partial\mathcal{L}^{\text{eff}}}{\partial\dot{t}} = \mu g_{00}^{\text{eff}} \dot{t}$ where $\dot\  = d/d\tau$.
Using the proper time definition, this leads to 
	\begin{equation}
	\begin{aligned}
	 \frac{k^2}{\left[ 1 - 2\frac{G_NM}{r}\left( 1 + 2\beta^2 - \lambda\beta^4G_NM^2  A \right) + \lambda\beta^4G_N^2 M^3 B\frac{\ln(\Lambda^2r^2)}{r} \right]}& - \left[ 1 + 2\frac{G_NM}{r}(1-2\beta^2)\right] \dot{r}^2 \\
	& - \frac{J^2}{\mu^2r^2\left[ 1 + 2\frac{G_NM}{r}(1-2\beta^2)\right]} = 1 .
	\end{aligned}
	\end{equation}
It is convenient to change  coordinate to $\tilde{r}^2=r^2\left[ 1 + 2\frac{G_NM}{r}(1-2\beta^2)\right]$, which corresponds to redefining $g_{\theta\theta}^{\text{eff}}=\tilde{r}^2$. At first order in $G_N$ we have $\dot{\tilde{r}}=\dot{r}$. Using the Binet variable $u=1/r$ such that $\dot{r}=-\frac{J}{\mu}\frac{du}{d\theta}$ we obtain
\begin{eqnarray}
&&\left(1-2G_N M(4\beta^2 -\lambda \beta^4 G_N M^2 A) u + {\lambda} \beta^4 G_N^2 M^3 B u \ln \frac{\Lambda^2}{u^2}\right)\left(\frac{du}{d\theta}\right)^2= \frac{(k^2-1)\mu^2}{J^2}\nonumber - u^2 \\ && \hspace{0,5cm} +\left(2G_N M(1+2\beta^2-\lambda\beta^4 G_N M^3 A) -\lambda \beta^4 G_N^2 M^3 B \ln \frac{\Lambda^2}{u^2}\right )u^3 \\ && \hspace{0,5cm} +\ \frac{2G_NM \mu^2 }{J^2}(1+2\beta^2 -\lambda \beta^4 G_N M^{2} A)u\nonumber -\frac{\mu^2}{J^2}\lambda \beta^4 G_N^2 M^3 {B} u\ln \frac{\Lambda^2}{u^2}.\\ \nonumber
\end{eqnarray}
Taking the first derivative of this identity, we finally obtain 
the Binet equation 
	\begin{equation}
	\begin{aligned}
	& \frac{d^2u}{d\theta^2}+u=\left( G_N\beta^2M(4-\lambda\beta^2G_NM^2(A-B)) - \lambda G_N^2\beta^4M^3B\ln(\frac{\Lambda}{u}) \right)\left(\frac{du}{d\theta}\right)^2 \\
	& \hspace{1cm} + \frac{G_N}{J^2}M\mu^2(1+2\beta^2-\lambda\beta^{4}G_NM^2(A-B)) - \frac{\lambda\beta^4G_N^2M^3\mu^2 B}{J^2}\ln(\frac{\Lambda}{u}) \\
	& \hspace{1cm} + G_NM(3-2\beta^2-\lambda\beta^4 G_NM^2(A-B))u^2 - \lambda\beta^4G_N^2 M^3Bu^2 \ln(\frac{\Lambda}{u})
	\end{aligned}
	\end{equation}
Solutions can be obtained  in perturbation theory around the classical trajectory $u=u_0+u_1$ where the Newtonian motion is given by 
\begin{equation}
	u_0 = \frac{G_NM\mu^2(1+2\beta^2-\lambda\beta^4G_NM^2(A-B))}{J^2} \left( 1+e\cos(\theta) \right) = \frac{1+e\cos(\theta)}{a(1-e^2)}
\end{equation}
where the semi long-axis is 
	\begin{equation}
	a=\frac{a_{\rm GR}}{(1+2\beta^2-\lambda\beta^{{4}}G_NM^2(A-B))},
\end{equation}
i.e. corrected compared to its GR value. 
The first order perturbation to the trajectory satisfies
	\begin{equation}
	\begin{aligned}
	& \frac{d^2u_1}{d\theta^2}+u_1=\left( G_N\beta^2M(4-\lambda\beta^2G_NM^2(A-B)) - \lambda G_N^2\beta^4M^3B\ln(\frac{\Lambda}{u_0}) \right)\left(\frac{du_0}{d\theta}\right)^2 \\
	& \hspace{0,5cm} - \frac{\lambda\beta^4G_N^2M^3\mu^2 B}{J^2}\ln(\frac{\Lambda}{u_0}) + G_NM\left(3-2\beta^2 -\lambda\beta^4G_NM^2(A-B)\right)u_0^2 - \lambda\beta^4 G_N^2M^3B\ln(\frac{\Lambda}{u_0})u_0^2 .
	\end{aligned}
	\end{equation}
The advance of  perihelion can be obtained by picking the  $\cos(\theta)$ in the source terms. The perturbation equation  becomes
\begin{equation}
	\begin{aligned}
	\frac{d^2u_1}{d\theta^2}+u_1 \supset & \left[ \lambda\beta^4G_N^2M^3B + {\lambda\beta^4G_NM^2B}{a(1-e^2)} + {2 G_NM}\left(3-2\beta^2+\lambda\beta^4G_NM^2(B-A)\right) \right. \\
	& \hspace{0,5cm} \left. + \lambda\beta^4G_N^2M^3B \left( 1 {-} 2\ln(\Lambda a(1-e^2)) \right) \right]\frac{e}{a^2(1-e^2)^2} \cos(\theta)	
	\end{aligned}
\end{equation}
whose solution is 
$
	u_1=e\alpha\theta\sin(\theta)
$
with
\begin{equation}
	\begin{aligned}
	\alpha = \left( 3-2\beta^2+\lambda\beta^4G_NM^2(2 B-A) \right)\frac{G_NM}{a^2(1-e^2)^2} + \frac{\lambda\beta^4G_NM^2B}{2 a(1-e^2)} {-} \lambda\beta^4G_N^2M^3B \frac{\ln(\Lambda a(1-e^2))}{a^2(1-e^2)^2} 
	\end{aligned}
\end{equation}
Thus we obtain at this order 
	\begin{equation}
	u= \frac{1+e\cos\big((1-\alpha a(1-e^2))\theta\big)}{a(1-e^2)}.
	\end{equation}
The advance of the perihelion is then given by
\begin{equation}
	\Delta w= 2\pi\alpha a(1-e^2).
\end{equation}	
The term in $3-2 \beta^2$ corresponds to the GR advance corrected by the direct coupling to matter, i.e it is the result for scalar-tensor theories of the Brans-Dicke type \cite{Will:2014kxa}. 

We can apply this result  to planets orbiting around the Sun like Mercury for which the contribution of GR to the perihelion advance is well known
	\begin{equation}
	\Delta w=w_{exp}-w_{GR}=0.0 {8}\pm 0.45
	\end{equation}
	in arc second per century \cite{Berche:2024cwe}. This gives us as upper bounds, $\beta\lesssim 10^{-1}$ corresponding to a maximal deviation to the GR result by around one percent and $\lambda\lesssim 10^{-72}$ when the Cassini bound $\beta^2 \lesssim 2.1 \times 10^{-5}$ is applied. The self coupling constant $\lambda$ should be very small. Although extremely small, such small couplings are also required in some dark matter model  and dark energy. We will discuss them in the conclusion. 
	
	\subsubsection{Shapiro time delay}
	
	The Shapiro effect is a gravitational time delay for  electromagnetic signals passing near a massive object.
	\begin{figure}[ht]
	\includegraphics[scale=0.55]{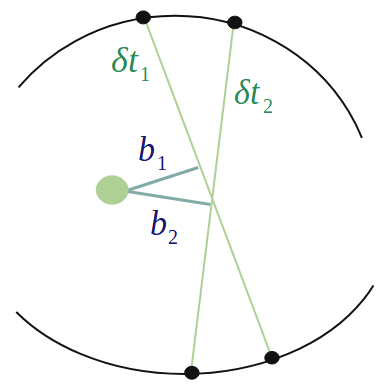}				
	\caption{Representation of Shapiro time delay for photon skirting a large objects with an impact parameter $b$}
	\end{figure}
 The supplementary time delay caused by the coupling of matter to a scalar field is due to the shift in the satellite's trajectories caused by the scalar forces. This entails a correction to the time taken by photons to reach the satellites. As such the effect can be calculated by taking null geodesics in the effective one body metric of the matter objects where we take $\nu=0$ to account for the mass difference between the central massive object and the very small satellites orbiting around it. In real experiments such as the measurements carried out by the Cassini probe \cite{Bertotti:2003rm}, the signals are emitted from Earth to the probe and back. The approximation $\nu\sim 0$ is  valid and one can consider that the photons travel along null geodesics of the effective one body metric whose external  gravitational field is due to the Sun.
 
We introduce the effective potential $\Phi(r)=\Phi_N(r)+\beta\frac{\varphi^0(r)}{m_{pl}}$, such that
	\begin{equation}
	\Phi(r)=-\frac{G_{\rm eff}M}{r} \hspace{0,5cm}\text{where}\hspace{0,5cm} G_{\rm eff}=(1+{2}\beta^2)G_N.
	\end{equation}
We consider at first order the photon trajectories to be straight lines with an impact parameter $b$. Thus in polar coordinates we have 
$
	r=\frac{b}{\cos(\theta)}.
$
We can the rewrite the effective one-body  metric as
\begin{equation}
	ds_{\rm eff}^2= -\left(1+2\Phi {+} 2\lambda\beta^4G_N^2\frac{M^3A\cos(\theta)}{b}+\lambda\beta^4G_N^2M^3B\frac{\ln(\Lambda^2r^2)\cos(\theta)}{b}\right)dt^2 + \left( 1 -2(1+2\gamma)\Phi \right)dx^2,
\end{equation}
where $dx^2=dr^2+r^2d\theta^2=\frac{b^2d\theta^2}{\cos^4(\theta)}$ along the trajectory and  
$
	\gamma=-\frac{2\beta^2}{1+2\beta^2}.
$
The photon trajectories are null  which gives then 
\begin{equation}
	\frac{dt}{dx} = 1 - 2(1+\gamma)\Phi {-} \cos(\theta) \frac{\lambda\beta^4G_N^2M^3}{b} \left( A {+} B \ln(\Lambda r) \right).
\end{equation}
The time delay due to the modification of gravity is
	\begin{equation}
	\frac{d\delta t}{dx} = 2\gamma\frac{G_{\rm eff}M}{b}\cos(\theta) {-} \cos(\theta) \frac{\lambda\beta^4G_N^2M^3}{b} \left( A {+} B \ln(\Lambda r) \right).
	\end{equation}
or in terms of the angle $\theta$ along the trajectory
\begin{equation}
	\frac{d\delta t}{d\theta} = 2\gamma\frac{G_{\rm eff}M}{\cos(\theta)} {-} \frac{\lambda\beta^4G_N^2M^3}{\cos(\theta)} \left( A {+} B \ln(\Lambda r) \right).
\end{equation}
Let us assume that the two  bodies where the signal is received and emitted follow circular trajectories,  in a first approximation, around another massive body. Thus $\cos(\theta_{e,r})=b/r_{e,r}$ and therefore a variation of the position of the emitter or the receiver by $d\theta_{e,r}$ corresponds to a change of impact parameter $db_{e,r} = -r_{e,r}\sin(\theta_{e,r})d\theta_{e,r}$. The total variation of the time delay due to a variation of the impact parameter is in this case 
\begin{equation}
	\frac{d\delta t}{db} = - \frac{M}{b}(2\gamma G_{\rm eff} {-} \lambda\beta^4G_N^2M^2A) \left( \frac{1}{\sin(\theta_e)} + \frac{1}{\sin(\theta_r)} \right) +\frac{\lambda\beta^4G_N^2M^3}{b} B \left( \frac{\ln(\Lambda r_e)}{\sin(\theta_e)} + \frac{\ln(\Lambda r_r)}{\sin(\theta_r)} \right).
\end{equation}
Measurement of this time delay are performed with probes such as Cassini \cite{Bertotti:2003rm} going behind the Sun for $\theta_{e,r}\sim\frac{\pi}{2}$. The time delay measured between two positions of the probe (emitter and receiver), after a round trip, with impact parameters $b_1$ and $b_2$ respectively is therefore
\begin{equation}
	\delta t_1 - \delta t_2 = \left[ - 8\gamma G_{eff}M {+} 4\lambda\beta^4G_N^2M^3A +2\lambda\beta^4G_N^2M^3B \ln(\Lambda^2 r_er_r) \right]\ln\left(\frac{b_1}{b_2}\right)
\end{equation}
which depends on the coupling and the  position of the emitter and receiver. The first term is the usual scalar-tensor result \cite{Will:2014kxa}.
The measurements by the Cassini probe give  experimentally \cite{Bertotti:2003rm} 
	\begin{equation}
	\gamma - 1 = (2.1 \pm 2.3)\times 10^{-5}
	\end{equation}
corresponding to  $\beta^2\lesssim 2\times 10^{-5}$ and the same bound on $\lambda$ as for the perihelion of Mercury, i.e  $\lambda\lesssim 10^{-72}$. All in all we can summarise the constraint from GR tests as
\be 
\lambda \beta^2\lesssim (G_N M^2_\odot)^{-1}
\ee
which we will discuss below.

\section{Radiation from the binary system}
\label{radiative}   

\subsection{The radiative effective action}    

Now that we have described the conservative action and the effects of the self-interaction on the binary system, we can move on to the radiative part. In this section we will set the scene and retrieve the well-known formula for the emitted power using  the Schwinger-Keldysh formalism. In the following section, we will derive  the tail effects.

The higher order terms in the action which mix the radiation and potential parts come from the Taylor expansion of the action (\ref{clfi}), see the appendix \ref{app:eff} for details,  $\delta S_1(\phi_1,\bar{\Phi}_{1,0}) = \delta S^{(2)} + \delta S^{(3)} + \delta S^{(4)} + \cdots$ where
\begin{equation}
    \begin{aligned}
    \delta S^{(2)} = \int d^4x \left( \frac{1}{2}\phi\square\phi - 6\lambda\bar{\Phi}^2\phi^2 - \frac{m^2}{2}\phi^2 \right), \hspace{0,4cm} \delta S^{(3)} = -\int d^4x\ 4\lambda\bar{\Phi}\phi^3, \hspace{0,4cm} \delta S^{(4)} = -\int d^4x \lambda\phi^4.
\end{aligned}
\end{equation}
We can recognise cubic and quartic interaction terms for the radiation field and a mass term coming from the interaction with the conservative sector. Formally and writing the effective propagator for the radiation field as a Dyson expansion, the mass insertion $12 \lambda \bar\Phi^2$ is eventually responsible for the tail effects affecting the motion of the binary system. We will expound this below.

Treating the case of a single field temporarily, the Klein-Gordon equation for the radiative $\phi$ is then 
\begin{equation}
    \square\phi= - \sum_n \frac{(-1)^n}{n!}\partial_{i_1}\cdots\partial_{i_n}\left(I^{i_1\cdots i_n}\delta^{(3)}(x)\right) + m^2\phi + 12\lambda\bar{\Phi}^2\phi + 12\lambda\bar{\Phi}\phi^2 + 4\lambda\phi^3 .
\end{equation}
The resulting action after substitution is then
\begin{equation}
    \delta S_{\rm eff}[x^{a,b}] = \int d^4x \sum_n \frac{1}{2\times n!}I^{i_1\cdots i_n}\delta^{(3)}(x)\partial_{i_1}\cdots\partial_{i_n}\bar \phi + 2\int d^4x \lambda\bar{\Phi}\bar \phi^3 + \int d^4x \lambda\bar \phi^4
\end{equation}
where $\bar\phi$ is the solution to the Klein-Gordon equation. This  gives the effective action for the binary system coming from the radiative part of the scalar field. In particular this will give rise to radiation-reaction forces. 

So far  we have not dealt with the two fields  in the Schwinger-Keldysh formalism. Let us reintroduce them.
It is convenient to change basis and consider  
\begin{equation}
    \phi_+ = \frac{\phi_1+\phi_2}{2} \hspace{0,5cm}\text{and}\hspace{0,5cm} \phi_-=\phi_1 -\phi_2.
\end{equation}
For example the source terms become  $I_+=\frac{I_1+I_2}{2}$ and $I_-=I_1-I_2$ such that
\begin{equation}
    \begin{aligned}
    \int d^4x \delta^{(3)}(x) & \left( \sum_n \frac{1}{n!}I_1^{i_1\cdots i_n}\partial_{i_1} \cdots \partial_{i_n}\phi_1 - \sum_n \frac{1}{n!}I_2^{i_1\cdots i_n}\partial_{i_1} \cdots \partial_{i_n}\phi_2 \right) \\
    & = \int d^4x \delta^{(3)}(x)\sum_n \frac{1}{n!} \left( I_-^{i_1\cdots i_n}\partial_{i_1} \cdots \partial_{i_n}\phi_+ + I_+^{i_1\cdots i_n}\partial_{i_1} \cdots \partial_{i_n}\phi_- \right) . 
    \end{aligned}
\end{equation}
In this new basis the action becomes
	\begin{equation}
	\begin{aligned}
	S^{rad}= & \int d^4x \left( \phi_-\square\phi_+ + \delta^{(3)}(x)\sum_n\frac{1}{n!}\left( I_-^{i_1\cdots i_n}\partial_{i_1}\cdots\partial_{i_n}\phi_+ + I_+^{i_1\cdots i_n}\partial_{i_1}\cdots\partial_{i_n}\phi_- \right) - m^2\phi_-\phi_+ \right. \\
	& \left. - 4\lambda\phi_-\phi_+(\phi_+^2+\frac{\phi_-^2}{4}) -4\lambda\phi_+^3(\Phi_1-\Phi_2) - 6\lambda\phi_-\phi_+^2(\Phi_1+\Phi_2) - 3\lambda\phi_-^2\phi_+(\Phi_1-\Phi_2) \right. \\
	& \left. - \frac{\lambda}{2}\phi_-^3(\Phi_1-\Phi_2) -6\lambda\phi_+^2(\Phi_1^2-\Phi_2^2) - 6\lambda\phi_+\phi_-(\Phi_1^2+\Phi_2^2) - \frac{3\lambda}{2}\phi_-^2(\Phi_1^2-\Phi_2^2) \right).
	\end{aligned}
	\end{equation}
The two Klein-Gordon  equations obtained from the radiative action are
\begin{equation}
    \begin{aligned}
    \square\phi_+ = -\sum_n & \frac{(-1)^n}{n!} \partial_{i_1}\cdots\partial_{i_n}\left(I_+^{i_1\cdots i_n}\delta^{(3)}(x)\right) + m^2\phi_+ + 4\lambda\phi_+^3 + 3\lambda\phi_+\phi_-^2 + 3\lambda\phi_-(\Phi_1^2-\Phi_2^2) \\
    & + 6\lambda\phi_+(\Phi_1^2+\Phi_2^2) + 6\lambda\phi_+^2(\Phi_1+\Phi_2) +6\lambda\phi_-\phi_+(\Phi_1-\Phi_2) + \frac{3}{2}\phi_-^2(\Phi_1-\Phi_2)
    \end{aligned}
\end{equation}
and 
\begin{equation}
    \begin{aligned}
    \square\phi_- = -\sum_n & \frac{(-1)^n}{n!} \partial_{i_1}\cdots\partial_{i_n}\left(I_-^{i_1\cdots i_n}\delta^{(3)}(x)\right) + m^2\phi_- + \lambda\phi_-^3 + 12\lambda\phi_+\phi_-^2 + 4\lambda\phi_-(\Phi_1^2+\Phi_2^2) \\
    & + 12\lambda\phi_+(\Phi_1^2-\Phi_2^2) + 12\lambda\phi_+^2(\Phi_1-\Phi_2) +12\lambda\phi_-\phi_+(\Phi_1+\Phi_2) + 3\phi_-^2(\Phi_1-\Phi_2).
    \end{aligned}
\end{equation}    
Replacing the solutions to the Klein-Gordon equation in the action, we obtain the effective action for the binary system in the radiative sector
\begin{equation}
	\begin{aligned}
	S_{\rm eff}^{\rm rad}[x^{a,b}_\pm] = & \int d^4x \left( \delta^{(3)}(x)\sum_n\frac{1}{n!} I_-^{i_1\cdots i_n}\partial_{i_1}\cdots\partial_{i_n}\bar\phi_+ +2\lambda\bar\phi_-^3 \bar\phi_+ -4\lambda\bar\phi_+^3(\bar\Phi_1-\bar\Phi_2) + 3\lambda\bar\phi_-^2\bar\phi_+(\bar\Phi_1-\bar\Phi_2) \right. \\
	& \hspace{2cm} \left. + \lambda\bar\phi_-^3(\bar\Phi_1-\bar\Phi_2) -6\lambda\bar\phi_+^2(\bar\Phi_1^2-\bar\Phi_2^2) + \frac{3\lambda}{2}\bar\phi_-^2(\bar\Phi_1^2-\bar\Phi_2^2) \right).
	\end{aligned}
	\end{equation}
 This action contains the dissipative and tail effects as we will see below.
\subsection{Dissipation}

Let us focus on the term coupling the sources to the radiative field $\int d^4x \delta^{(3)}(x) I_-^{i_1\cdots i_n}\partial_{i_1}\cdots\partial_{i_n}\bar\phi_+$. Taking $\bar \phi_+$ at the lowest order and using the matrix of propagators in the $(\phi_+,\phi_-)$ basis \cite{Goldberger:2004jt,Porto:2016pyg,Jakobsen:2022psy}\footnote{The $++$ element of the matrix corresponds to the Hadamard propagator which is the sum of the positive and negative frequency propagators with no action on the quasi-static classical sources and therefore is dropped here, see \cite{Galley:2009px}.}
	\begin{equation}
	G_{ab}= 	\begin{pmatrix}
0 & G_{adv}\\
G_{\rm ret} & 0
\end{pmatrix}
	\end{equation}
where
\begin{equation}
	G_{\rm ret}(t-t', x-x')=\frac{1}{4\pi}\frac{\delta(t-t'+|x-x'|)}{t-t'}
	\end{equation}
we find  at leading order
\begin{equation}
	\begin{aligned}
	\phi_+^{(0)} & =-\sum_n\frac{(-1)^n}{n!} \int d^4x' G_{\rm ret}(x-x') \partial_{i_1}\cdots\partial_{i_n}\left(I_+^{i_1\cdots i_n}\delta^{(3)}(x')\right)  \\ & = -\sum_n\frac{1}{n!} \int dt'G_{\rm ret}(\vec x,t-t')n_{i_1}\cdots n_{i_n}\partial_0^nI_+^{i_1\cdots i_n}
 \label{solu}
	\end{aligned}
	\end{equation}
using that $\partial_iG_{\rm ret}(x-x')=-n_i\partial_0G_{\rm ret}(x-x')$ where $n_i$ points between $x'$ and $x$. Explicitly we have then 
\begin{equation}
	\begin{aligned}
	\sum_n \frac{1}{n!} & \int d^4x \delta^{(3)}(x)I_-^{i_1\cdots i_n}\partial_{i_1}\cdots \partial_{i_n}\phi^{(0)}_+ \\
	& = -\sum_{n,n'} \frac{1}{n!n'!}\int dtdt'd^3x \delta^{(3)}(x)I_-^{i_1\cdots i_n}n_{i_1}\cdots n_{i_n}n_{j_1}\cdots n_{j_{n'}} G_{\rm ret}(\vec x,t-t') \partial_0^n I_-^{i_1\cdots i_n}\partial_0^{n'} I_+^{j_1\cdots j_{n'}} \\
	& = -\sum_{n} \frac{1}{n!(2n+1)!!}\int dtdt' \left( \partial_0^n I_-^{i_1\cdots i_n}\right) G_{\rm ret}(\vec 0, t-t') \partial_0^{n'} \left( I_+^{j_1\cdots j_{n'}} \right)
	\end{aligned}
	\end{equation}
The last integral is obtained as  ($n^A=n^i\cdots n^j$)
	\begin{equation}
	\int d^3x\delta^{(3)}(x)n^A G_{\rm ret}(\vec x,t-t')=\frac{1}{4\pi}\int d\Omega \vert \vec x \vert ^2d\vert \vec x \vert \frac{\delta(\vert \vec x\vert)}{\vert \vec x\vert ^2}n^A=\frac{1}{4\pi}\int d\Omega n^A=<n^A>
	\end{equation}
 where , see appendix \ref{d} for details,
\begin{equation}
	<n_{i_1}\cdots n_{i_n}n_{j_1}\cdots n_{j_{n'}}>=\frac{4\pi}{(n+n'+1)!!}\left( \delta_{i_1i_2}\cdots\delta_{j_{n'-1}j_{n'}} + \text{symmetry}\right).
\end{equation}
As the multipoles  are trace free we have $n=n'$ and  $n!$ contractions.
As a result  we obtain the dissipative action for the binary system
	\begin{equation}
	S_{\rm diss}[x_{\pm}^{a,b}]\equiv \sum_n \frac{1}{n!} \int d^4x \delta^{(3)}(x)I_-^{i_1\cdots i_n}\partial_{i_1}\cdots \partial_{i_n}\phi^{(0)}_+ = -\frac{1}{4\pi}\sum_n \frac{1}{(2n+1)!!n!}\int dt \left(\partial_0^n I_-^{i_1\cdots i_n}\right)\left(\partial_0^{n+1} I_+^{i_1\cdots i_n}\right) .
	\end{equation}
where we have used\footnote{ This follows from  the fact that for any distribution $f$ we have  $\int dt'G_{\rm ret}(\vec 0, t-t')f(t')=\int \slashed{d}\omega f(\omega)\int {\slashed{d}^3k}\frac{1}{(\omega-i\epsilon)^2-k^2-m^2}$. The last integral does not make sense and we regularise it as $m\rightarrow 0$ by introducing $\alpha\ge 0$
	\begin{equation}
	\begin{aligned}
	\lim_{m\to 0}\int \frac{d^3k}{(2\pi)^3}\frac{1}{(\omega-i\epsilon)^2-k^2-m^2}  & = \frac{4\pi}{(2\pi)^3} \int_0^{\infty} \frac{k^2dk}{(\omega-i\epsilon)^2-k^2}  \underset{\alpha\rightarrow 0}{=}  \frac{2\pi}{(2\pi)^3} \int_{\mathbb{R}} \frac{k^2dk}{(\omega-i\epsilon)^2-k^2} e^{i\alpha k} \\
	& = \frac{2i\pi}{(2\pi)^2}\left[ \frac{k^2}{-2k} \right]_{-(\omega-i\epsilon)} = \frac{i\omega}{4\pi}
	\end{aligned}
	\end{equation}}
	\begin{equation}
	\int dt'G_{\rm ret}(\vec 0,t-t')f(t')=\int d\omega\frac{i\omega}{4\pi}e^{i\omega t}f(\omega)=\frac{1}{4\pi}\frac{d}{dt}f(t).
	\end{equation}
The force acting on the binary system due to dissipation can be obtained as
\begin{equation}
	\frac{\delta S_{diss}}{\delta x_-} = -\frac{1}{4\pi}\sum_n \frac{1}{(2n+1)!!n!} \frac{\delta}{\delta x_-} \left[\left(\partial_0^n I_-^{i_1\cdots i_n}\right)\left(\partial_0^{n+1} I_+^{i_1\cdots i_n}\right)\right] .
\end{equation}
When $x_-=0$, we have $I_+=I$ and then we obtain  $\left\lbrace \begin{array}{l}
	\frac{\delta I_+}{\delta x_-}=0 \\
	\frac{\delta I_-}{\delta x_-}= \frac{\delta I}{\delta x}
\end{array}\right.$. This implies that the equations of motion simplify to 
	\begin{equation}
	m_{a,b}\ddot{x}_{a,b}+\frac{\partial V}{\partial x_{a,b}} = -\frac{1}{4\pi}\sum_n \frac{1}{(2n+1)!!n!} \frac{\delta}{\delta x_{a,b}} \left(\partial_0^n I^{i_1\cdots i_n}\right)\partial_0^{n+1} I^{i_1\cdots i_n}
	\end{equation}
 where the total variation only acts on the bracketed term. 
This is Newton's law in the presence of a radiation-reaction force coming from the emission of scalars by the binary system. Multiplying by $\dot{x}_{a,b}$ we get using $\sum_{i=a,b}\dot{x}_{i}\frac{\delta f(x)}{\delta x_{i}}=\partial_0 f(x)$ the energy balance equation
	\begin{equation}
	\begin{aligned}
	\frac{d}{dt}\left( \frac{m_{a}}{2}\dot{x}_{a}^2+\frac{m_{b}}{2}\dot{x}_{b}^2+V \right) = -\frac{1}{4\pi}\sum_n \frac{1}{(2n+1)!!n!} \left(\partial_0^{n+1} I^{i_1\cdots i_n}\right)\partial_0^{n+1} I^{i_1\cdots i_n}.
	\end{aligned}
	\end{equation}
This gives us the time variation of the conservative energy as $\frac{dE}{dt}=P$ with~\cite{Ross:2012fc,Porto:2016pyg,Kuntz:2019zef}
	\begin{equation}
	P= -\frac{1}{4\pi}\sum_n \frac{1}{(2n+1)!!n!} \left(\partial_0^{n+1} I^{i_1\cdots i_n}\right)^2 
	\end{equation}
The power $P$ is negative by construction and this coincides with (\ref{flux}) when the radiative solution (\ref{solu}) is used \cite{Ross:2012fc}.
Notice that the previous derivation of the balance equation between the radiated scalar energy by the binary system and the variation of its conservative energy has been derived using a Lagrangian approach despite the presence of dissipation. 
	   
\subsection{The multipole expansion}
    
The multipole expansion depends on the source term ${\cal J}$ which contains two parts. The first one comes from the coupling to matter and can be explicitly calculated using the matter action. The leading terms are given by ${\cal J}_{\rm matter}=-\bar J$  where we have \cite{Brax:2019tcy} 
\begin{eqnarray}
m_{\rm Pl}{\cal J}_{\rm matter}&=& -\beta m_a \delta^{(3)}(x-x_a)- \beta m_b\delta^{(3)}(x-x_b)+  \beta ( m_a v_a^2 \delta^{(3)}(x-x_a)\nonumber \\ &&+ m_b v_b^2 \delta^{(3)}(x-x_b)) - \frac{\beta G_N(1+2\beta^2)m_a m_a}{r}(\delta^{(3)}(x-x_a)+\delta^{(3)}(x-x_b))\nonumber \\
\end{eqnarray}
where $r= \vert \vec x_a- \vec x_b\vert$. In the centre of mass frame, this leads to the first few multipoles \cite{Brax:2019tcy}
\begin{eqnarray}
    I_m &=& -\frac{\beta M}{m_{\rm Pl}} + \frac{8\beta G_N m_A m_B}{3m_{\rm Pl}r}
    \label{eq:Iphi}
    \nonumber \\
    I^i_m &=& 0
   \nonumber \\
    I^{ij}_m &=& -\frac{\beta \mu}{m_{\rm Pl}} \left(r^i r^j - \frac{\delta^{ij}}{3} r^2\right).\nonumber \\
   \label{eq:Iphiij}
\end{eqnarray}
The vanishing of the dipole comes from the condition $m_a \vec x_a + m_b \vec x_b=\vec 0$ in the centre of mass frame. This is peculiar to the choice of universal coupling $\beta$ of the scalar field to matter. The higher order multipoles can be obtained similarly\footnote{
We will be interested in the dominant part of the emitted power 
    \begin{equation}
    P=-\frac{1}{4\pi} \left( (\dot{I})^2 + \frac{1}{3}(\ddot{I^i})^2 + \frac{1}{30}(\dddot{I^{ij}})^2 \right).
    \end{equation}
which we will compare to the usual gravitational emission.
In the scalar case without self-interactions, the dipole contribution vanishes. The monopole and quadrupole terms are of the same order of magnitude whilst the higher multipole terms are suppressed by $v^{2(\ell -2)}$ for the multipoles of order $\ell$.   }. The new ingredient for self-interacting scalars is the source provided by action of the conservative sector on radiation. This follows from
\be 
{\cal J}^\lambda=-4 \lambda (\bar \Phi^{(0)})^3.
\ee
Contrary to the point particle nature of the sourcing by matter, i.e. involving Dirac distributions, this source is distributed in space and time.
\begin{equation}
    \begin{aligned}
    {\cal J}^{\lambda} & = 4\lambda \int d^4y d^4z d^4w\int \slashed{d}^4 p \slashed{d}^4q \slashed{d}^4 l \frac{e^{-ip_0(x_0-y_0)}e^{i\vec{p}(\vec{x}-\vec{y})}}{\vec{p}^2-p_0^2+i\epsilon} \frac{e^{-iq_0(x_0-y_0)}e^{i\vec{q}(\vec{x}-\vec{y})}}{\vec{q}^2-q_0^2+i\epsilon} \frac{e^{-il_0(x_0-y_0)}e^{i\vec{l}(\vec{x}-\vec{y})}}{\vec{l}^2-l_0^2+i\epsilon} \\
    & \hspace{2cm} \times \frac{1}{M_{\varphi}^3}\big(m_a\delta(\vec{y}-\vec{x_a}(t))+m_b\delta(\vec{y}-\vec{x_b}(t))\big)\big(m_a\delta(\vec{z}-\vec{x_a}(t))+m_b\delta(\vec{z}-\vec{x_b}(t))\big) \\ & \hspace{4,5cm} \big(m_a\delta(\vec{w}-\vec{x_a}(t))+m_b\delta(\vec{w}-\vec{x_b}(t))\big).
    \end{aligned}
\end{equation}
Notice that we have used the Feynman propagators. 
As before, we only consider the leading terms in $p_0/|\vec{p}|\ll 1$ and the terms mixing the sources as they are the only ones affecting the dynamics of the system. This gives rise to terms of the following type (and the one obtained by exchanging $a$ and $b$)
    \begin{equation}
    {\cal J}^{\lambda}(t,\vec{x}) = 4\lambda\frac{m_am_b^2}{M_{\varphi}^3} \int \slashed{d}^3 p \slashed{d}^3q \slashed{d}^3 l  \frac{e^{i\vec{p}(\vec{x}-\vec{x}_a)+i(\vec{q}+\vec{k})(\vec{x}-\vec{x}_b)}}{\vec{p}^2\vec{q}^2\vec{l}^2}.
    \end{equation}
It is convenient to use the Fourier representation of the multipoles    
\begin{equation}
    {\cal J}^{\lambda}(t,\vec{k}) = 4\lambda\frac{m_am_b^2}{M_{\varphi}^3} e^{-i\vec{k}.\vec{x}_b} \int \slashed{d}^3 p \slashed{d}^3q   \frac{e^{i\vec{p}(\vec{x_a}-\vec{x_b})}}{\vec{p}^2\vec{q}^2(\vec{p}+\vec{q}+\vec{k})^2}.
\end{equation}
We consider soft external momenta  such $\vec{|k|} \ll \vec{|p|},\vec{|q|}$.
Writing the square $(\vec{p}+\vec{q}+\vec{k})^2 = (\vec{p}+\vec{q})^2+2(\vec{p}+\vec{q}).\vec{k}+\vec{k}^2$, we expand the propagator as    
\begin{equation}
    \frac{1}{(\vec{p}+\vec{q}+\vec{k})^2}=\frac{1}{(\vec{p}+\vec{q})^2}\Big(1-2\frac{(\vec{p}+\vec{q}).\vec{k}}{(\vec{p}+\vec{q})^2} - \frac{\vec{k}^2}{(\vec{p}+\vec{q})^2}+4\frac{(\vec{p}.\vec{k})^2+(\vec{q}.\vec{k})^2}{(\vec{p}+\vec{q})^4}+8\frac{(\vec{p}.\vec{k})(\vec{q}.\vec{k})}{(\vec{p}+\vec{q})^4}+ \cdots \Big).
\end{equation}
Additionally we expand $e^{-i\vec{k}.\vec{x}_b}=1-i\vec{k}.\vec{x}_b-\frac{(\vec{k}.\vec{x}_b)^2}{2} + \cdots$ implying that     
    \begin{equation}
    \begin{aligned}
    {\cal J}^{\lambda}(t,\vec{k}) & = 4\lambda\frac{m_am_b^2}{M_{\varphi}^3} \Big[ \left(1-i\vec{k}.\vec{x}_b -\frac{(\vec{k}.\vec{x}_b)^2}{2}\right) \int \slashed{d}^3 p \slashed{d}^3q  \frac{e^{i\vec{p}.\vec{r}}}{\vec{p}^2\vec{q}^2(\vec{p}+\vec{q})^2}\\ & - 2\left(1-i\vec{k}.\vec{x}_b\right)\int \slashed{d}^3 p \slashed{d}^3q \frac{e^{i\vec{p}.\vec{r}}(\vec{p}.\vec{k}+\vec{q}.\vec{k})}{\vec{p}^2\vec{q}^2(\vec{p}+\vec{q})^4}  + 4\int \slashed{d}^3 p \slashed{d}^3q  \frac{e^{i\vec{p}.\vec{r}}}{\vec{p}^2\vec{q}^2(\vec{p}+\vec{q})^6} \left((\vec{p}.\vec{k})^2+(\vec{q}.\vec{k})^2+2(\vec{p}.\vec{k})(\vec{q}.\vec{k}) \right)\\ & -\int \slashed{d}^3 p \slashed{d}^3q  \frac{e^{i\vec{p}.\vec{r}}}{\vec{p}^2\vec{q}^2(\vec{p}+\vec{q})^4}\vec{k}^2 \Big].
    \end{aligned}
    \end{equation}    
This  gives explicitly the monopole, dipole and quadrupole contributions as
\begin{equation}
    \begin{aligned}
    & {\cal J}_{\lambda}(t) = 4\lambda\frac{m_am_b^2}{M_{\varphi}^3} \int\slashed{d}^3 p \slashed{d}^3q  \frac{e^{i\vec{p}.\vec{r}}}{\vec{p}^2\vec{q}^2(\vec{p}+\vec{q})^2} \\
    & {\cal J}_{\lambda}^i(t) =  4\lambda\frac{m_am_b^2}{M_{\varphi}^3} \Big[ x_b^i \int \slashed{d}^3 p \slashed{d}^3q  \frac{e^{i\vec{p}.\vec{r}}}{\vec{p}^2\vec{q}^2(\vec{p}+\vec{q})^2} -2i \int  \slashed{d}^3 p \slashed{d}^3q \frac{e^{i\vec{p}.\vec{r}}}{\vec{p}^2\vec{q}^2(\vec{p}+\vec{q})^4}(p^i+q^i) \Big] \\
    & {\cal J}_{\lambda}^{ij}(t) = 4\lambda\frac{m_am_b^2}{M_{\varphi}^3} \Big[ x_b^ix_b^j \int \slashed{d}^3 p \slashed{d}^3q  \frac{e^{i\vec{p}.\vec{r}}}{\vec{p}^2\vec{q}^2(\vec{p}+\vec{q})^2} -4ix_b^i \int \slashed{d}^3 p \slashed{d}^3q  \frac{e^{i\vec{p}.\vec{r}}}{\vec{p}^2\vec{q}^2(\vec{p}+\vec{q})^4}(p^j+q^j)\\ &  \hspace{1,3cm} + 2\int \slashed{d}^3 p \slashed{d}^3q  \frac{e^{i\vec{p}.\vec{r}}}{\vec{p}^2\vec{q}^2(\vec{p}+\vec{q})^4} \delta^{ij}  - 8\int \slashed{d}^3 p \slashed{d}^3q  \frac{e^{i\vec{p}.\vec{r}}}{\vec{p}^2\vec{q}^2(\vec{p}+\vec{q})^6}(p^ip^j+q^iq^j+2p^iq^j) \Big].
    \end{aligned}
\end{equation}    
These integrals may diverge for small momenta. We will introduce the mass $m$ of the scalar field  to regularise the IR divergences and later take the limit $m\rightarrow 0$ at the end of the calculation. We will see that the physical observables do not suffer from IR issues. 
All the above integrals can be obtained from the template
\begin{equation} \label{m}
    \int\slashed{d}^3 p \slashed{d}^3q   \frac{e^{i\vec{p.r}}}{(\vec{p}^2+m^2)(\vec{q}^2+m^2)}\frac{(p^{\alpha_1}+q^{\alpha_1})...(p^{\alpha_D}+q^{\alpha_D})}{((\vec{p}+\vec{q})^2+m^2)^n} = (-i)^D\partial^{\alpha_1}...\partial^{\alpha_D} K_n(\alpha)\vert_{\alpha=0}
\end{equation}
where the generating function is given by
\be
    K_n(\alpha) = \int \slashed{d}^3 p \slashed{d}^3 u \frac{e^{i\vec{p}.\vec{r}+i\vec{u}.\vec{\alpha}}}{(\vec{p}^2+m^2)((\vec{u}-\vec{p})^2+m^2)}\frac{1}{(u^2+m^2)^n}
\ee
We compute $K_n(\alpha)$ in the appendix \ref{a} and obtain the explicit representation
\begin{equation}
    \begin{aligned}
    & K_n(\alpha)=\frac{i}{16\pi^2}\int_0^1dx \sum_{k=0}^{\infty}(-1)^k\frac{|\vec{\alpha}+\vec{r}(1-x)|^{2k}}{(2k+1)!}\frac{m^{2(k-n)+2}}{(n-1)!}A(n,k,x) \\
    & A(n,k,x)=\frac{d^{n-1}}{dv^{n-1}}\left( \frac{v^{2k+2}}{(v+i)^n \sqrt{1+x(1-x)v^2}} \right) \biggr\arrowvert_{v=i}
    \end{aligned}
\end{equation}
Finally, we use Mathematica to compute the integrals $A(n,k,x)$. Putting everything together, i.e. each contribution mixing the sources, we get for the monopole, dipole and quadrupole terms
\begin{equation}
    \begin{aligned}
    & {\cal J}_{\lambda}(t) = -\lambda\frac{m_am_b}{M_{\varphi}^3} \frac{3\log(3)}{8\pi^2}(m_a+m_b) \\
    & {\cal J}_{\lambda}^i(t) =  -\lambda \frac{m_am_b}{M_{\varphi}^3} \Big[ \frac{3\log(3)}{8\pi^2}(m_ax_a^i+m_bx_b^i) + \frac{(2+\log(27))}{24\pi^2}(m_b-m_a)r^i \Big] \\
    & {\cal J}_{\lambda}^{ij}(t) = -\lambda \frac{m_am_b}{M_{\varphi}^3} \Big[ \frac{3\log(3)}{8\pi^2}(m_ax_a^ix_a^j+m_bx_b^ix_b^j) + \frac{(2+\log(27))}{{12}\pi^2}(m_bx_b^i-m_ax_a^i)r^j -\frac{m^{-2}}{2\pi^2}(m_a+m_b)\delta^{ij} \\ & \hspace{0,5cm}- \frac{(4+9\log(3))}{48(2\pi)^2}(m_a+m_b)r^2\delta^{ij} + \frac{11m^{-2}}{18\pi^2}(m_a+m_b)\delta^{ij} +\frac{(26/9+\log(27))}{20(2\pi)^2}(m_a+m_b)\left(r^2\delta^{ij}+2r^ir^j\right) \Big]
    \end{aligned}
\end{equation}
The terms of the form $constant\times\delta^{ij}$ in the quadrupole disappear from the effective action since they give total derivative terms as $\delta^{ij} \partial_i\partial_j\varphi = \triangle\varphi = \ddot{\varphi}$ using the wave equation. 
The dipole simplifies using the centre of mass condition $({m}_a\vec{x}_a+{m}_b\vec{x}_b)$  which we set to 0.
We can  write the quadrupole in terms of $Q^{ij}=(r^ir^j-\frac{1}{3}r^2\delta^{ij})$ as
\begin{equation}
    \begin{aligned}
    {\cal J}_{\lambda}^{ij}(t) = -\lambda \frac{\beta^3{\mu M^2}}{8\pi m_{\rm Pl}^3}\left[ C Q^{ij} + D r^2\delta^{ij} \right] 
    \end{aligned}
\end{equation}
with
    \begin{equation}
    \begin{aligned}
    & C =  \left[ \frac{26/9+\log(27)}{5\pi}+ \left( \frac{3\log(3)}{\pi} - 2\frac{(2+\log(27))}{3\pi} \right)\nu \right] \\
    & D =  \left[ \left( \frac{(26/9+\log(27))}{6\pi} - \frac{(4+9\log(3))}{24\pi} \right) + \frac{1}{3} \left( \frac{3\log(3)}{\pi} - 2\frac{(2+\log(27))}{3\pi} \right)\nu \right].
    \end{aligned}
    \end{equation}
The terms in the quadrupole that are proportional to $\delta^{ij}$ can be inserted in the monopole using the wave equation and a double integration by parts. So we get the multipoles,
    \begin{equation}
    \begin{aligned}
    & I_{\lambda}(t) = -\frac{\lambda\beta^3 D \mu M^2}{8\pi m^3_{\rm Pl}}  \ddot{(r^2)} ,\\
    & I_{\lambda}^i(t) =  -\frac{\lambda\beta^3 m_am_b}{m_{\rm Pl}^3}  \frac{(2+\log(27))}{24\pi^2}(m_b-m_a)r^i   ,\\
    & I_{\lambda}^{ij}(t) = -\frac{\lambda\beta^3 C\mu M^{2}}{8\pi m_{\rm Pl}^3} Q^{ij} .
    \end{aligned}
    \end{equation}
As advocated, the result is finite when the mass is taken to vanish $m\to 0$ with no IR divergences. We can now consider the effects of the scalar multipoles on the emitted power.  

\subsection{The scalar emitted power}

Before calculating the emitted power from scalars, let us recall that the power emitted in the form of gravitational waves is 
\be 
P_{\rm GR}=-\frac{G_N \mu^2}{5}\left\langle \left( \dddot{Q^{ij}} \right)^2 \right\rangle.
\ee
The quadrupole contribution just adds to the graviton power and we have
    \begin{equation}
    P^{(2)}=  \left( 1 + \frac{\beta^2}{3}(1 +\lambda \beta^2C{G_N M^2})^2  \right) P_{\rm GR}
    \end{equation}
The emitted power in GR can be expressed in terms of
\be 
x=\frac{G_N M}{a} = (G_N M \omega)^{2/3}
\ee
with $\omega = \frac{2\pi}{T}$ where  $T$ is the period of the orbit satisfying Kepler's third law $\omega^2a^3= (G_N M)^{-1} $ and is given by
\be 
P_{\rm GR}=-\frac{32}{5}\frac{\nu^2}{G_N} \frac{1+ \frac{73}{24}e^2+ \frac{37}{96}e^4}{(1-e^2)^{7/2}}x^5
\ee
when averaged over one orbital period. This will be useful when comparing with the power emitted by scalars in the monopole. 

First of all, the dipole contribution to the emitted power comes from the self-interactions only and is therefore in $\lambda^2$, i.e. higher order in $\lambda$, and can therefore be discarded.
The complete monopole for the scalar field is given by 
 \begin{equation}
     I = -\frac{\lambda\beta^3 D\mu M^2}{8\pi m^3_{\rm Pl}}  \ddot{(r^2)}  + \frac{8\beta G_N\mu M}{3m_{\rm Pl}r}
 \end{equation}
 and so 
 \begin{equation}
     \left\langle \left( \dot{I} \right)^2 \right\rangle = 8\pi \lambda^2\beta^6D^2 \mu^2 M^4 G_N^3 \left\langle  \dddot{\left(r^2\right)} ^2 \right\rangle + \frac{512 \pi\beta^2G_N^3\mu^2M^2}{9} \left\langle \dot{\left(\frac{1}{r}\right)} ^2 \right\rangle - \frac{128 \pi\lambda \beta^4 G_N^3 D\mu^2 M^3}{3} \left\langle \dddot{\left(r^2\right)} \dot{\left(\frac{1}{r}\right)} \right\rangle .
 \end{equation}
 We obtain using
 \be 
 \langle f(t)\rangle= \frac{(1-e^2)^{3/2}}{2\pi}\int_0^{2\pi}\frac{d\theta}{(1+e\cos \theta)^2} f(t(\theta))
 \ee
and $dt= \frac{a^2(1-e^2)^2}{L(1+e\cos\theta)^2} d\theta$ that the required averages are simply
\begin{equation}
    \begin{aligned}
    & \left\langle \dddot{(r^2)}^2 \right\rangle = L^6\frac{e^2(4+e^2)}{2a^8(1-e^2)^{13/2}} = \frac{e^2(4+e^2)}{2G_N^2{M}^2(1-e^2)^{7/2}}{x}^5 \\
    & \left\langle \dot{\left(\frac{1}{r}\right)}^2 \right\rangle = L^2\frac{e^2(4+e^2)}{8a^6(1-e^2)^{9/2}} = \frac{e^2(4+e^2)}{8G_N^4{M}^4(1-e^2)^{7/2}}{x}^5 \\
    & \left\langle \dddot{\left(r^2\right)} \dot{\left(\frac{1}{r}\right)} \right\rangle = L^4\frac{e^2(4+e^2)}{4a^7(1-e^2)^{11/2}} = \frac{e^2(4+e^2)}{4(1-e^2)^{7/2} G_N^3 M^4}{x}^5
    \end{aligned}
 \end{equation}
 so we can add these contributions to the power and get
 \be 
 P^{(0)}=- \frac{e^2(4+e^2)}{(1-e^2)^{7/2}} \left(1+ \frac{9\lambda^2 \beta^4 D^2 G_N^2 M^4}{8}- \frac{3\lambda\beta^2 D G_N M^2}{2} \right)\frac{16 \beta^2 \nu^2}{9G_N} x^5
 \ee
This vanishes when $e=0$ and otherwise is proportional to the GR case. As a result the effects of the scalar field do not give corrections to the frequency dependence of gravitational waves. On the other hand the slight increase of power due to the scalar in the quadrupole emission could be constrained with pulsar timing \cite{Benisty:2022lox}. 
Notice that the corrections to the case of a simple conformal coupling case modifying  GR at the $\beta^2$ order appears in the form of $\lambda \beta^2 G_N M^2$. This is the same combination and therefore the same order of magnitude as a constraint on $\lambda$  as obtained for the advance of perihelion and the Shapiro effect.

	\section{Tail effects}
 \label{tail}
We have only calculated one term in the binary system effective action as calculated in the Schwinger-Keldysh formalism. This has allowed us to retrieve the known expression for the power emitted in the scalar sector. We will not compute all the terms in the action here  and focus on one effect which can be deduced directly, i.e. the one leading to tail terms. Other terms and their physical relevance are left for future work.

\subsection{Conservative-radiative coupling}

The tail effects  are corrections to the radiation-reaction of the binary system which involve the memory of the past trajectory of the binary system. 
	\begin{figure}[ht]
	\includegraphics[scale=0.4]{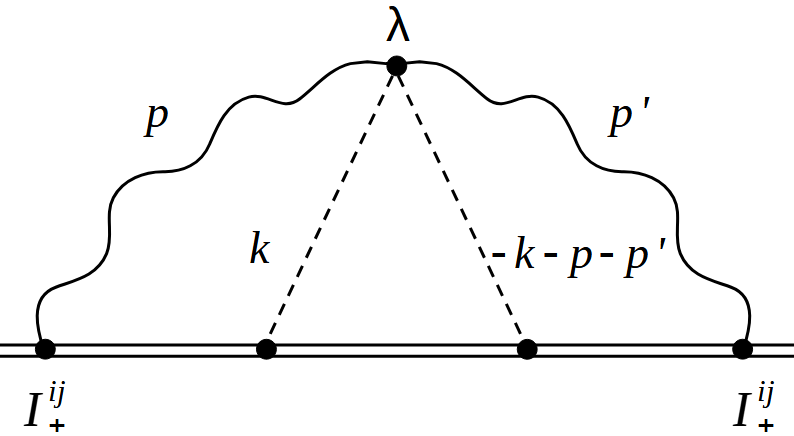}				
	\caption{The diagram associated to the scalar tail effect. There are two radiation fields (wiggly line) coupled to the $I_+$ source and two conservative fields (dotted lines).}
	\end{figure}	
	In the effective action there are terms coupling radiative and potential fields. We will focus on one class of term which will give a non-vanishing result in the physical limit when $x_-=0$ and $\phi_-=0$. The term $-6\lambda\int d^4x \left(\Phi_1^2-\Phi_2^2\right)\phi_+^2$ can be evaluated in several steps. Let us concentrate on 
\begin{equation}
	 \int d^4x \Phi_1^2\phi_+^2 =  \int d^4x \Phi_1^2 \left( \int d^4x G_{ret}(x-x')J_+(x') \right)^2.
\end{equation}
This involves an integral which eventually must be multiplied by $-6 \lambda \frac{\beta^2m_am_b}{m_{\rm Pl}^2}$ as we are only interested in terms involving two different sources in $\Phi_1^2$
\be 
\int \slashed{d}^3 p \slashed{d}^3p'\slashed{d}^3 k \slashed{d}^3 q \slashed{d}\omega \frac{\slashed\delta^{(3)}(\vec{k}+\vec{q}+\vec{p}+\vec{p'}) J_+(p)J_+(p')e^{i\vec{k}.\vec{x}_{1,b}}e^{i\vec{q}.\vec{x}_{1,a}}}{(\vec{k}^2+m^2-i\epsilon)(\vec{q}^2+m^2-i\epsilon)((\omega+i\epsilon)^2-\vec{p}^2-m^2)((-\omega+i\epsilon)^2-\vec{p'}^2-m^2)}  
\ee
This can be expanded as 
\be 
\sum_{n,n'}\frac{1}{n!n'!}\int\slashed{d}^3 p \slashed{d}^3p'\slashed{d}^3 q \slashed{d}\omega   \frac{ I_+^{i_1\cdots i_n}(\omega)(-ip_{i_1})\cdots(-ip_{i_n}) I_+^{j_1\cdots j_{n'}}(-\omega)(-ip_{j_1})\cdots(-ip_{j_{n'}})e^{i\vec{k}.\vec{x}_{1,b}}e^{i\vec{q}.\vec{x}_{1,a}}}{((\vec{q}+\vec{p}+\vec{p'})^2+m^2-i\epsilon)(\vec{q}^2+m^2-i\epsilon)((\omega+i\epsilon)^2-\vec{p}^2-m^2)((-\omega+i\epsilon)^2-\vec{p'}^2-m^2)} \\
	\hspace{4cm} \times \slashed\delta^{(3)}(\vec{k}+\vec{q}+\vec{p}+\vec{p'})
\ee
since the Fourier transform of the source term reads $J_+(p)=\sum_n\frac{1}{n!}I_+^{i_1\cdots i_n}(\omega)(-ip_{i_1})\cdots(-ip_{i_n})$.
This will be given by 
	\begin{equation}
\partial_{\lambda^{i_1}}\cdots\partial_{\lambda^{i_n}}\partial_{\lambda^{j_1}}\cdots\partial_{\lambda^{j_{n'}}} \left( \int \slashed{d} \omega I_+^{i_1\cdots i_n}(\omega) I_+^{j_1\cdots j_{n'}}(-\omega) J(\omega,\lambda,\lambda') \right)\biggr\vert_{\lambda=\lambda'=0}
	\end{equation}
where we introduce the generating functional 
	\begin{eqnarray}
	&& J(\omega,\lambda,\lambda')=\nonumber \\
 && \hspace{0,3cm} \int\slashed{d}^3 p \slashed{d}^3p' \slashed{d}^3 q  \frac{ e^{-i\vec{\lambda}.\vec{p}-i\vec{\lambda'}.\vec{p'}} e^{-i(\vec{p}+\vec{p}'+\vec{q}).\vec{x}_{1,b}}e^{i\vec{q}.\vec{x}_{1,a}}}{((\vec{q}+\vec{p}+\vec{p'})^2+m^2-i\epsilon)(\vec{q}^2+m^2-i\epsilon)((\omega+i\epsilon)^2-\vec{p}^2-m^2)((-\omega+i\epsilon)^2-\vec{p'}^2-m^2)}. \nonumber \\
	\end{eqnarray}
The details of the computation can be found in the appendix \ref{c}. We obtain
	\begin{equation}
	J(\omega,\lambda,\lambda')= \frac{1}{16(2\pi)^4i}(I-I^*) 
 \end{equation}
where we have introduced 
\be 
 I=\int_0^1dx\int_0^1dy\frac{1}{\sqrt{x(1-x)y(1-y)}}\frac{\left(K(i\beta-\alpha)+\frac{i\pi}{2}H_0^{(1)}(\beta+i\alpha)\right)}{|\vec{X}|},
 \ee
 where $ \vec{X}=(1-x)\vec{\lambda'}+x\vec{\lambda}+(1-y)\vec{x}_{1,b}+y\vec{x}_{1,a}$ and $\beta=\frac{\omega X}{\sqrt{x(1-x)}}, \ \alpha=\omega |\vec{x}_{1,b}-\vec{x}_{1,a}|\sqrt{\frac{y(1-y)}{x(1-x)}}$.
The integrals are obtained as a function of 
	\begin{equation}
	K(x)=-i\int_0^{\pi/2}d\theta e^{x\cos(\theta)} \hspace{0,5cm}\text{and}\hspace{0,5cm} H_0^{(1)}(x)=\frac{2}{i\pi}\int_0^{\infty}dz e^{ix\cosh(z)}.
	\end{equation}
 the latter $H_0^{(1)}$ being the  Hankel function of the first kind of zeroth order.
	
As we are interested in the tail effects, they depend on the past of the system. The memory effects only appear when $\ln \abs{\omega}$ or inverse powers of $\omega$ arise in the series expansion of the functions $K$ and $H_0^{(1)}$. These terms  only appear in the logarithms characteristic of the Hankel functions of zeroth order. As a result we focus on the terms
	\begin{equation}
\partial_{\lambda^{i_1}}\cdots\partial_{\lambda^{i_n}}\partial_{\lambda^{j_1}}\cdots\partial_{\lambda^{j_{n'}}} \left( \int\slashed{d}\omega I_+^{i_1\cdots i_n}(\omega) I_+^{j_1\cdots j_{n'}}(-\omega)\frac{\left(H_0^{(1)}(\beta+i\alpha)+H_0^{(1)}(\beta+i\alpha)^*\right)}{X} \right)\biggr\vert_{\lambda=\lambda'=0}
	\end{equation}
We will first consider the lower order terms and then give the general result for the scalar tail effects. 	

\subsection{Monopole-monopole term}	
 For the the monopole-monopole term, i.e. $n=n'=0$, we use the series expansion of the Hankel functions truncated at the lowest order. More details about this expansion can be found in appendix (\ref{hankel})
\begin{equation}
	\begin{aligned}
	H_0^{(1)}(z) \approx & 1+\frac{2i}{\pi}\left(\gamma-\ln(2)+\ln(z)\right) - \frac{i}{2\pi}\left(\gamma-1-\frac{i\pi}{2}-\ln(2)+\ln(z)\right)z^2 \\
	& \hspace{2cm} + \frac{i}{32\pi}\left(\gamma-\frac{3}{2}-\frac{i\pi}{2}-\ln(2)+\ln(z)\right)z^4  + \mathcal{O}(z^6)
	\end{aligned}
	\end{equation}
with $z=\beta+i\alpha=\vert \omega\vert T \Delta$ , {{ we can write $\ln(z)=\ln(|\omega|T )+\ln(\Delta)$ where $T$ is the period of the binary motion}}\footnote{The choice of the period $T$ in the logarithmic term is arbitrary \cite{Bernard:2016wrg}, see appendix \ref{app:mem}. This choice will reappear in the calculation of the periastron.}. Keeping only terms in $\ln(|\omega|)$ since the others will not give memory effects
	\begin{equation}, 
	H_0^{(1)}(z)+H_0^{(1)}(z)^* \approx  \frac{2}{\pi}\alpha\beta\ln(\vert \omega\vert T) - \frac{1}{4\pi}\beta\alpha(\beta^2-\alpha^2)\ln(\vert \omega\vert T) + \cdots
	\end{equation}	
	We have $\beta= \omega\frac{X}{\sqrt{x(1-x)}}$ and $\alpha= \omega r_1\sqrt{\frac{y(1-y)}{x(1-x)}}$. The index $r_1$ refers to the first branch on the Keldysh contour. The term of the form $\omega^2\ln(|\omega|)$ will be the first one leading to  memory effects. In the  monopole-monopole case  the tail effect is given by, see the appendix \ref{app:mem} for details,  
	\begin{equation}
	\begin{aligned}
	& -\frac{3\lambda \beta^2 \mu M}{8(2\pi)^4 m_{\rm Pl}^2} \int_0^1dx\int_0^1dy \frac{\sqrt{y(1-y)}/x(1-x)}{\sqrt{x(1-x)y(1-y)}}  \int dt r_1\dot{I}_+(t) \int_{-\infty}^{t} dt' \frac{\dot{I}_+(t')}{(t-t')} \\
	& \hspace{3cm} = -\frac{3\lambda \beta^2\mu M}{m^2_{\rm Pl}} \frac{B(-1/2,-1/2)}{8(2\pi)^4} \int dt r_1 \dot{I}_+(t) \int_{-\infty}^{t} dt' \frac{\dot{I}_+(t')}{(t-t')}
	\end{aligned}
	\end{equation}
	where $B$ is the Euler Beta function which is extended analytically to the complex plane except for negative integers. We find that the first derivative of the monopoles are coupled via the same non-local Kernel in $\theta(t)/t$ as in the GR case. This pattern will reappear at higher order.  We will also find that selection rules prevent some multipoles from interacting. 

The previous results apply to the integrals $\int d^4x \Phi_2^2\phi_+^2$ too. These tail terms in the action lead to radiation-reaction forces. These forces are  obtained  by taking the derivative of the term $-6\lambda\int d^4x \left(\Phi_1^2-\Phi_2^2\right)\phi_+^2$ with respect to $x_{-}$ in the physical limit ($x_-=0$). Since $\frac{\delta I_+}{\delta x_-}=0$, the derivative only acts on the part in $r$. For  the dominant monopole-monopole  interaction we find
	\begin{equation}
	\begin{aligned}
	& \frac{\delta}{\delta x_{-,a}^k}\left( r_1-r_2 \right) = \frac{\delta x_{1,a}^l}{\delta x_{-,a}^k}\frac{\delta r_1}{\delta x_{1,a}^l} - \frac{\delta x_{2,a}^l}{\delta x_{-,a}^k}\frac{\delta r_2}{\delta x_{2,a}^l}= \frac{n_{1,k}}{2}+\frac{n_{2,k}}{2}
	\end{aligned}
	\end{equation}
	writing $\vec{r}=r\vec{n}$. The derivative with respect to $x_{-,b}$ gives the opposite sign.  We obtain  the force exerted on the $a^{th}$ body from the monopole-monopole term as
	\begin{equation}
	\begin{aligned}
	& F_{k,a}= \frac{\delta}{\delta x_{-,a}^k}\left[-\frac{6\lambda\beta^2 \mu M}{m^2_{\rm Pl}}\int d^4x \left(\Phi_1^2-\Phi_2^2\right)\phi_+^2\right]\biggr\vert_{x_-=0} \supset -\frac{3\lambda\beta^2\mu M n_k}{8(2\pi)^4 m_{\rm Pl}^2} B(-1/2,-1/2) \dot{I}_+(t) \int_{-\infty}^{t} dt' \frac{\dot{I}_+(t')}{(t-t')} .
	\end{aligned}
	\end{equation}
	The power correcting the balance equation is $\vec{F}_a.\vec v_a+ \vec F_b\vec{v}_b= \vec F_a .\vec v$ where $\vec v= \vec v_a -\vec v_b$. For the monopole-monopole term, this  gives 
	\begin{equation}
	P_{\rm tail}=\vec{F}_a.\vec{v} \supset -\frac{3\lambda \beta^2 \mu M \vec n .\vec v}{8(2\pi)^4 m^2_{\rm Pl}} B(-1/2,-1/2) \dot{I}_+(t) \int_{-\infty}^{t} dt' \frac{\dot{I}_+(t')}{(t-t')} 
	\end{equation}
	We will evaluate it for closed Newtonian orbits below.

	\subsection{Monopole-dipole term}	
	
The next interesting case is the monopole-dipole term which corresponds to  $n=1$, $n'=0$ or $n'=1$, $n=0$. The $\lambda$ and $\lambda'$ dependence only appears in  $X$  which comes from $\beta$. The first order contribution in $\omega$ to the memory effects is the one in $\beta^3\alpha\ln(|\omega|T )$ and so in $\omega^4\ln(|\omega| T)$ leading to 
	\begin{equation}
	\begin{aligned}
	& \partial_{\lambda^i}\left(\frac{H_0^{(1)}(z)+H_0^{(1)}(z)^*}{X}\right) = \partial_{\lambda^i}\left( -\frac{X^2r_1\sqrt{y(1-y)}}{4\pi(x(1-x))^{2}}\omega^4\ln(|\omega|T) \right)
	\end{aligned}
	\end{equation}
Using the fact that we work in the centre of mass frame, we have 
\begin{equation}
	\partial_{\lambda^i}\left(X^2\right)\vert_{\lambda=0}=2x(y-\frac{m_a}{m_a+m_b})r_{1,i} \hspace{0,5cm}\text{and}\hspace{0,5cm} \partial_{\lambda'^j}\left(X^2\right)\vert_{\lambda=0,\lambda'=0}=2(1-x)(y-\frac{m_a}{m_a+m_b})r_{1,j}.
	\end{equation}
	So the monopole-dipole tail effect is
	\begin{eqnarray}
	&& \frac{3\lambda \beta^2 \mu M }{32(2\pi)^4{m^2_{\rm Pl}}} B(-1/2,-3/2) \int_0^1dy (y-\frac{m_a}{m_a+m_b}) \left( \int dt r_1^2 \ddot{I}_+^i(t)n_{1,i} \int_{-\infty}^{t} dt' \frac{\ddot{I}_+(t')}{(t-t')} \nonumber \right. \\   && \hspace{1cm} \left. + \int dt r_1^2\ddot{I}_+(t) \int_{-\infty}^{t} dt' \frac{\ddot{I}_+^j(t')n_{1,j}}{(t-t')} \right)
	\nonumber \\
	\end{eqnarray}
	at the lowest order. The integral over $y$ can be calculated and we find
 \begin{equation}
	\begin{aligned}
	& \frac{3\lambda \beta^2 \mu M }{64(2\pi)^4m^2_{\rm Pl}} B(-1/2,-3/2) \frac{m_b-m_a}{m_a+m_b} \left( \int dt r_1^2 \ddot{I}_+^i(t)n_{1,i} \int_{-\infty}^{t} dt' \frac{\ddot{I}_+(t')}{(t-t')} + \int dt r_1^2 \ddot{I}_+(t) \int_{-\infty}^{t} dt' \frac{\ddot{I}_+^j(t')n_{1,j}}{(t-t')} \right)
	\end{aligned}
	\end{equation}
    which involves the second derivatives of the monopole and the dipole. Notice that the same non-local kernel as in the monopole-monopole case reappears. 
We will generalise this result below.

\subsection{Higher order tail interactions}
We use the asymptotic expansion of the Hankel function of zeroth order, see appendix (\ref{hankel}), evaluated for  $z=\beta+i\alpha=\vert \omega\vert (b+ia)$. This gives 
    \begin{equation}
    H_0^{(1)}(z)\simeq \frac{2i}{\pi}\sum_{m\geq0}\frac{(-1)^m\omega^{2m}}{(m!)^22^{2m}}(b+ia)^{2m}\ln(|\omega| T) .
    \end{equation}
We now have to evaluate $H_0^{(1)}+(H_0^{(1)})^*$ using
\begin{equation}
    \begin{aligned}
    i(b+ia)^{2m}-i(b-ia)^{2m} & = i\sum_{p=0}^{2m}C_{2m}^p (b^{2m-p}(ia)^p-b^{2m-p}(-ia)^p) = -2 \sum_{p'=0}^{m-1}(-1)^{p'} C_{2m}^{2p'+1} b^{2m-2p'-1}a^{2p'+1}
    \end{aligned}
\end{equation}
so we get
    \begin{equation}
  H_0^{(1)}(z)+H_0^{(1)}(z)^*\simeq    -\frac{4}{\pi}\ln(|\omega| T) \sum_{m\geq1} (-1)^m\frac{\omega^{2m}}{m!^22^{2m}}\sum_{p'=0}^{m-1}(-1)^{p'} C_{2m}^{2p'+1} b^{2m-2p'-1}a^{2p'+1}
    \end{equation}
    and dividing by X before taking derivatives gives us
    \begin{equation}
      -\frac{4}{\pi}\ln(|\omega|T) \sum_{m\geq1} (-1)^m\frac{\omega^{2m}}{m!^22^{2m}}\sum_{p'=0}^{m-1}(-1)^{p'} C_{2m}^{2p'+1} X^{2m-2p'-2}r^{2p'+1}(x(1-x))^{-m}(y(1-y))^{\frac{1}{2}(2p'+1)} .
    \end{equation}
    This can be used to derive the multipole-multipole interactions. 
    \subsubsection{The monopole-monopole}
    For the particular case of monopole-monopole we evaluate the previous expression for $\lambda=\lambda'=0$ and use 
    $X\equiv X_0=r|y-\frac{m_a}{M}|$. We must also perform the integration over $x$ and $y$. This gives
    \begin{equation}
    \int_{[0,1]^2}dxdy \left(-\frac{4}{\pi}\right)\ln(|\omega| T)\sum_{m\geq1}(-1)^m\frac{\omega^{2m}}{(m!)^22^{2m}}\sum_{p'=0}^{m-1}(-1)^{p'} C_{2m}^{2p'+1} X^{2m-2p'-2}r^{2p'+1} (x(1-x))^{-m-\frac{1}{2}}(y(1-y))^{p'}
    \end{equation}
    Putting it all together, we find the general expression for the monopole-monopole interaction
    \begin{equation}
    \begin{aligned}
    & \frac{3\lambda \beta^2 \mu M }{4(2\pi)^4 m^2_{\rm Pl}}\sum_{m\geq1}\frac{(-1)^m}{(m!)^22^{2m}} B(-m+1/2,-m+1/2) \sum_{p'=0}^{m-1}(-1)^{p'} C_{2m}^{2p'+1} F(m_a,m_b,p') \\
    & \hspace{5cm}\int dt r^{2m-1}  I^{(m)}(t)\int_{-\infty}^t dt' \frac{I^{(m)}(t)}{t-t'}
    \end{aligned}
    \end{equation}
	where $I^{(m)}(t)= \frac{d^m}{dt^m} I(t)$ and $F(m_a,m_b,p')=\int_0^1 dy |y-\frac{m_a}{M}|^{2m-2p'-2}(y(1-y))^{p'}$.
 This generalises the case obtained previously corresponding to $m=1$. The terms of higher order involve derivatives of order $m$ in time. 
 
\subsubsection{The higher multipole interactions}
    For the higher order terms, we must act with derivatives with respect to $\lambda$ or $\lambda'$. We remark that for  $p'=n-1$ we have a term in $X^0$ so no derivatives can act on it. This implies that we must consider derivatives 
    \begin{equation}
    \partial_{\lambda}^n\partial_{\lambda'}^{n'}\left(-\frac{4}{\pi}\ln(|\omega|T)\sum_{n\geq2}\frac{\omega^{2m}}{(n!)^22^{2m}}(x(1-x))^{-m}(-1)^m\sum_{p'=0}^{m-2} (-1)^{p'}C_{2m}^{2p'+1}X^{2m-2p'-2}r^{2p'+1}(y(1-y))^{\frac{1}{2}(2p'+1)}  \right)
    \end{equation}
    where, writing $R=(1-y)yx_b+yx_a$, we have
    \begin{equation}
    \begin{aligned}
    & X=(1-x)\lambda'+x\lambda +R \\
    & X^2 = ((1-x)\lambda'+x\lambda)^2 + 2R.((1-x)\lambda'+x\lambda) + R^2
    \end{aligned}
    \end{equation}
    As the derivatives act on  the term $(X^2)^{m-p'-1}$, we find a selection rule $n+n'\leq m-p'-1$  and the explicit expressions
    \begin{equation}
    \partial_\lambda^n\partial_{\lambda'}^{n'} (X^2)^{m-p'-1} \biggr\vert_0 = 2^{n+n'}x^n(1-x)^{n'}\vec{R}^{n+n'}(X_0^2)^{m-p'-1-n-n'}(m-p'-1)\cdots(m-p'-1-n-n'+1) 
    \end{equation}
    where $X_0^2=|R|^2=(y-\frac{m_a}{M})^2r^2$. Putting everything together we find that the derivatives of $(H_0+H_0^*)/X$ are given by 
    \begin{equation}
    \begin{aligned}
    & - \frac{4}{\pi}\ln(|\omega|T) \vec{r}^{N+N'} \sum_{m\geq2} \frac{\omega^{2m}}{(m!)^22^{2m}}x^{n'-m}(1-x)^{n'-m}(-1)^m r^{2m-1} \sum_{p'=0}^{m-1-n-n'} (-1)^{p'}C_{2m}^{2p'+1} \\ 
    & \hspace{3cm} \times\left(y-\frac{m_a}{M}\right)^{2(m-p'-1)} (y(1-y))^{p'+\frac{1}{2}} (m-p'-1)\cdots(m-p'-n-n').
    \end{aligned}
    \end{equation}
    This implies that the multipole-multipole interactions in the binary system action read
    \begin{equation}
    \begin{aligned}
    & \frac{3\lambda \beta^2 \mu M}{4(2\pi)^4m^2_{\rm Pl}}  \sum_{m\geq2} \frac{(-1)^m2^{n+n'}}{n!n'!(m!)^22^{2m}} B(n-m+1/2,n'-m+1/2) \sum_{p'=0}^{m-1-n-n'} (-1)^{p'} C_{2m}^{2p'+1} F(m_a,m_b,p') \\ 
    & \hspace{1cm} \times  \frac{(m-p'-1)!}{(m-p'-n-n'-1)!}  \int dt r^{2m-1-n-n'} I_+^{(m),i_1\cdots i_n}n_{i_1}\cdots n_{i_n} \int_{-\infty}^t \frac{dt}{t-t'} I_+^{(m),i_1\cdots i_{n'}}n_{i_1}\cdots n_{i_{n'}}. \\
    \end{aligned}
    \end{equation}
	for a given order of interaction between $n$-poles and $n'$-poles. For the monopole-dipole we have $n=0, n'=1$ or $n=1, n'=0$ whose lowest order starts at $m=2$. Notice that the structure is similar as the one for the monopoles. We also find that the result is extremely similar to the one in GR where the quadrupoles are coupled with the characteristic tail effect involving the convolution with the kernel in $\theta(t)/t$. Finally, we have selection rules appearing in the sum over $p'$ and leading to interactions between multipoles only when $n+n'+1\le m$ where $m$ is the order of the time derivatives acting on the multipoles. We retrieve that the monopole-monopole term involves first derivatives at leading order, the monopole-dipole second derivatives and the monopole-quadrupole contains third order time derivatives.

 \subsection{Monopole-monopole effects on the phase of gravitational waves}

Let us now give an estimate of the orders of magnitude for the various multipole-multipole interactions. As for the case of the monopole-monopole interaction, there is an associated power contributing to the balance equation of the binary system. At the lowest order we simply evaluate the power coming from the tail terms in the Newtonian approximation of the closed trajectories.
The next term is the monopole-dipole interaction which vanishes as the dipole only receives a contribution at order $\lambda$ and is zero at the Newtonian order. The next term would be the monopole-quadrupole interaction which involves third time derivatives and is reduced by a factor of $v^4$ compared to the monopole-monopole case.
In fact all the higher order terms come with extra powers of $a/T$ as each extra time derivative on the multipoles is accompanied by a factor of $r$. Now using Kepler's third law we have $a/T\sim (G_N M/a)^{1/2} \sim v$ implying that they are all higher order in the Post-Newtonian expansion. As a result, the monopole-monopole is the dominant term. 

We are interested in the average over the orbits
\begin{equation}
\langle P_{\rm tail}\rangle = -\frac{3\lambda \beta^2 \mu M }{8(2\pi)^4 m^2_{\rm Pl}} B(-1/2,-1/2) \langle (\vec n.\vec v)\dot{I}_+(t) \int_{-\infty}^{t} dt' \frac{\dot{I}_+(t')}{(t-t')} \rangle.
	\end{equation}
This can be evaluated by using the Fourier series \cite{Bernard:2016wrg,Trestini:2024mfs}
\be 
I_+(t)= \sum_n I_n e^{2\pi int/T}
\ee
as $I_+(t)$  is periodic of period $T$. 
In practice and at leading order the contribution from the conformal coupling is 
\be 
I_n= \frac{8G_N \beta M\mu}{3m_{\rm Pl}}\frac{1}{T}\int_{-\frac{T}{2}}^{\frac{T}{2}} \frac{dt}{r}e^{-2\pi int/T}
\ee
which is given by the real terms $I_{-n}= \bar I_n= I_n$.
It is useful to utilise the representation $r= a (1-e \cos \eta)$ and $t= \frac{T}{2\pi} (\eta -e\sin \eta)$. This gives
\be 
I_n=\frac{8G_N \beta M\mu}{3m_{\rm Pl}}\frac{1}{2\pi a}\int_{-\pi}^{\pi}  d\eta e^{-in(\eta- e \sin \eta)}= 
\frac{8G_N \beta M\mu}{3m_{\rm Pl}a} J_n(en)
\ee
in terms of Bessel functions.
It is then convenient to write the real Fourier series as 
\be 
I_+(t)= I_0 + 2 \sum_{n>0} I_n \cos \left(2\pi n\frac{t}{T}\right).
\ee
Using this we define 
\begin{eqnarray}
&&{\cal P}=\langle (\vec n.\vec v)\dot{I}_+(t) \int_{-\infty}^{t} dt' \frac{\dot{I}_+(t')}{(t-t')} \rangle \nonumber \\ && \hspace{0,4cm} = \sum_{n>0,n'>0} \frac{2(2\pi)^2 nn'}{T^2} I_n I_{n'}\langle  {\sin(2\pi n\frac{t}{T})}(\vec v. \vec n) \int_{-\infty}^{+\infty}\frac{dt' \theta(t-t')}{i(t-t')} (e^{i2\pi n't'/T} - e^{-i2\pi n' t'/T})\rangle \nonumber \\
\end{eqnarray}
which becomes 
\begin{eqnarray} 
&&{\cal P}=\sum_{n,n'} \frac{4(2\pi)^2nn'}{T^2}\ln(2\pi \vert  n'\vert ) I_n I_{n'}\langle  {\sin(2\pi n\frac{t}{T})\sin(2\pi n'\frac{t}{T})}(\vec v.\vec n)\rangle \nonumber \\ && \hspace{0,4cm}=\sum_{n,n'} \frac{2(2\pi)^2 nn'}{T^2}\ln(2 \pi\vert  n')\vert  I_n I_{n'}  \int_{-\frac{T}{2}}^{\frac{T}{2}} dt (\cos(2\pi (n-n')\frac{t}{T}) - \cos(2\pi (n+n')\frac{t}{T}))(\vec v. \vec n).
\nonumber \\
\end{eqnarray}
 In this case we have 
\begin{eqnarray}
&&\int_{-\frac{T}{2}}^{\frac{T}{2}} dt (\cos(2\pi (n-n')\frac{t}{T}) - \cos(2\pi (n+n')\frac{t}{T}))(\vec v. \vec n)\nonumber \\ && = e  \int_{-\pi}^{\pi} d\eta \sin \eta (\cos((n-n')(\eta-e\sin\eta)) {-} \cos ((n+n')(\eta-e\sin\eta)))=0\nonumber \\
\end{eqnarray}
As a result the tail effect coming from the monopole-monopole does not have any secular consequences on the size of the orbits. The same result can be extended to all multipole-multipole interactions as the integrand always picks a $\sin \eta$ from the $\vec v . \vec n$ term whilst the rest of the integrand is an even function of $\eta$. In fact the only secular effects can be obtained in the form of the periastron advance as we will see below.

\subsection{Advance of Periastron}

Evaluated at leading order along Newtonian orbits, the monopole-monopole part of the tail effects is a  radial force such that its contribution to the acceleration in the centre of mass is 
\be 
\vec a \supset {\cal R} \vec n
\ee
where we have denoted by ${\cal R}$ the radial component of the force 
\be 
{\cal R}=  -\frac{3\lambda \beta^2  M }{8(2\pi)^4 m^2_{\rm Pl}} B(-1/2,-1/2) \dot{I}_+(t) \int_{-\infty}^{t} dt' \frac{\dot{I}_+(t')}{(t-t')} 
\ee
which reads explicitly
\be 
{\cal R}=-\frac{3\lambda \beta^2  M }{8(2\pi)^2 m^2_{\rm Pl}} B(-1/2,-1/2)\sum_{n>0,n'>0} \frac{2nn'}{T^2}\ln (2\pi\vert  n'\vert ) I_n I_{n'}   (\cos(2\pi (n-n')\frac{t}{T}) {-} \cos \left(2\pi (n+n')\frac{t}{T}\right)) 
\ee
When the orbits are parameterised by 
$ r=\frac{p}{1+e\cos \theta }, \dot{r} =\sqrt{\frac{Gb{M}}{p}} e\sin \theta ,
r\dot{\theta} = \sqrt{\frac{Gb{M}}{p}}(1+e\cos \theta)
$ where $b=1+2\beta^2$ and $p=a(1-e^2)$, 
this  leads to the periastron advance \cite{poisson_will_2014}
\begin{eqnarray}
    \frac{dw}{d{\theta}} \simeq \frac{1}{e} \frac{p^2}{Gb{M}} \left[ - \frac{\cos \theta }{(1+e\cos \theta)^2} {\cal R} + \frac{2+e\cos \theta}{(1+e\cos \theta)^3}\sin \theta {\cal  S} \right]
\end{eqnarray}
as a function of the radial ${\cal R}$ and orthogonal ${\cal S}$ components of the acceleration. Here ${\cal S}=0$ as the force is radial.
Similarly the eccentricity and the periastron evolve according to 
\begin{eqnarray}
    \frac{dp}{d\theta} & \simeq& 2 \frac{p^3}{Gb{M}} \frac{\cal S}{(1+e\cos \theta )^3} \equiv 0 
    \label{eq:pevoln}
    \\
     \frac{de}{d\theta} &\simeq & \frac{p^2}{Gb{M}} \left[  \frac{\sin \theta}{(1+\cos \theta )^2} {\cal R} + \frac{2\cos \theta +e(1+\cos^2 \theta)}{(1+e\cos \theta)^3} {\cal S} \right]
\end{eqnarray}
where we retrieve the absence of secular effects on the size of the orbits.  It is more convenient to use the $\eta$ representation with 
\be 
\tan \frac{\theta}{2}= \sqrt{\frac{1+e}{1-e}}\tan \frac{\eta}{2}
\ee
from which
\be 
\cos \theta= \frac{\cos \eta -e}{1-e\cos \eta}.
\ee
This allows us to obtain 
\be 
\frac{dw}{d\eta}=\frac{a^{{2}}\sqrt{1-e^2}}{e{ GbM}}(e-\cos \eta){\cal R}
\ee
Integrating over the orbit we obtain the total advance of periastron
\be 
\Delta \omega= -\frac{3\lambda \beta^2 a^{{2}}\sqrt{1-e^2} }{{2} e {b}} B(-1/2,-1/2)\sum_{n>0,n'>0} \frac{2nn'}{T^2}\ln (2\pi \vert  n'\vert ) I_n I_{n'} J_{n,n'}  
\ee
where
\be 
J_{n,n'}=\frac{1}{2\pi}\int_{-\pi}^\pi d\eta(e- \cos\eta) (\cos((n-n')(\eta-e\sin \eta)) {-} \cos ((n+n')(\eta -e \sin\eta) ))
\ee
is given by
\begin{eqnarray}
&&J_{n,n'}=e( J_{(n-n')}\left({(n-n')} e\right) - J_{(n+n')}\left({(n+n')}e\right))\nonumber \\ &&-\frac{1}{2}( J_{(n-n')+1}\left((n-n')e\right)+ J_{(n-n')-1}\left((n-n')e\right) -J_{{(n+n')}+1}\left({(n+n')}e\right)-J_{{(n+n')}-1}\left((n+n')e\right)).
\nonumber
\end{eqnarray}
This can be summarised as
\be 
\Delta w= -\frac{\lambda\beta^2 (1-e^2)^{3/2}}{{2} (2\pi)^{{3}}}B(-1/2,-1/2) C(e) \Delta w_{GR}
\ee
where 
\be 
C(e)= \frac{1}{e}\sum_{n>0,n'>0} {nn'}\ln (2\pi\vert  n'\vert ) I_n I_{n'} J_{n,n'}
\ee
which has a finite limit when $e\to 0$. A similar calculation shows that the variation of the eccentricity vanishes by parity. 

As an order of magnitude we have  that
\be 
C(e) = {\cal O}(\beta^2 \left(\frac{G_N M}{a}\right)^2 G_N \mu^2 )
\ee leading to
\be
\frac{\Delta w}{\Delta w_{\rm GR}}= {\cal O}(\lambda \beta^4 \left(\frac{G_N M}{a}\right)^2 G_N \mu^2 )
\ee
For Mercury whose semi major axis is of  $58$ million km and its mass is 6 million smaller than a solar mass, we have then $G_NM/a\simeq 2\times 10^{-8}$ and $\mu/m_{\rm Pl}\sim 10^{32}$. Imposing that $\Delta w/\Delta w_{GR} \lesssim 10^{-2}$ we find that $\lambda \beta^4 \lesssim  10^{-50} $ which is weaker than the bounds we obtained already. We can also consider the S2 stars around Sagittarius ${\rm A}^*$ whose Schwarzschild radius is $1.27 \times 10^{10}$ m and has a  semi major axis of $a= 1.53\times 10^{14}$ m corresponding to $G_N M/a \simeq 10^{-4}$, see \cite{Tedesco:2023sey} for instance. Using the bound $ \lambda \beta^2 G_N M_\odot^2 \lesssim 1$ we have $\frac{\Delta w}{{\Delta w}_{GR}} \simeq \lambda \beta^4(G_N M/a)^2 G_N M_\odot^2 \lesssim \beta^2 (G_N M/a)^2\ll 1$, so even in this case the periastron advance is minimal. Larger effects would require enhanced values of $G_N M/a$ corresponding to nearly relativistic effects and a significant  value of $\beta$ which would bypass the solar system constrained. This might be achievable in certain scalarisation models \cite{Ramazanoglu:2016kul} where a small value of $\beta$ in the solar system could be compatible with a large coupling $\beta$ to binary systems. We leave this to further study. 

 \section{Discussion and Conclusion}

Light scalar fields with self-interactions could play a major role in cosmology. They could comprise some part of the dark matter of the Universe and alleviate some of the small scale issues of the $\Lambda$-CDM model. They could also be the local manifestation of dark energy fields coupled to matter. In this article, we presented the consequences of quartic self-interactions for (nearly) massless  fields mediating a long range force between massive bodies in the Universe. We consider the effects in the conservative sector of binary systems such as the advance of perihelion of Mercury and the Shapiro effect. We find that the self-coupling $\lambda$ is tightly constrained and must satisfy $\lambda \beta^2 G_N M^2_\odot \lesssim 1$ where the Yukawa coupling to matter $\beta^2\lesssim 2\times 10^{-5}$ according to the Cassini measurement. In the radiative sector we give a new derivation for the emitted power by scalar fields and consider tail effects. We perform these calculations in the Schwinger-Keldysh formalism as it naturally incorporates the conservative and dissipative effects. The tail effect have the same type of memory  as in GR and involve interactions between all the multipoles of the binary system. We find that they lead to no secular drift of the orbital size. On the other hand they induce a small advance of the periastron which would be maximal for very massive bodies. Very massive black holes would be such candidates although their direct coupling $\beta$ to the scalar field would be depleted by the no hair theorems. We have left phenomenological implications and limitations for further study.

Very large self-interactions for massive scalars representing dark matter are excluded as the typical self-interaction cross section $\sigma \simeq \frac{\lambda^2}{m^2}$ of these scalars would be too large for events like the bullet cluster where two large galaxy clusters collide. Typically this leads to the upper bound $\lambda \lesssim 10^{-12} (\frac{m}{\rm eV})^{3/2}$ \cite{Brax:2019fzb} which is more lax for $m\lesssim 10^{-20}$ eV associated with the size of the solar system than the bound we have obtained. More studies on the phenomenology of these self-interacting dark matter models with a direct coupling to matter are left for future work.  For models of dark energy one can estimate dimensionally that $\lambda \simeq \frac{V_0}{m^4_{\rm Pl}} \simeq 10^{-120}$ where $V_0$ is the amount of dark energy. This is much smaller than the bounds we obtained leading to no effects from such dark energy scalars. Dynamical dark energy, if existing, would involve time-dependent field configurations, it would be interesting to study the gravitational impact of such a slow time dependence. This is left for future work. 
\acknowledgments

We would like to thank L. Bernard, A. Kuntz and P. Vanhove for crucial remarks. E.B thanks IPhT for hospitality.

\appendix

\section{Integrating out short distance fields}
\label{app:eff}
The short and large distance fields are defined by having supports in Fourier space defined by $\vert \vec p \vert \ge \frac{1}{L}$ and $\vert \vec p \vert \le \frac{1}{L}$ where $L$ is the length separating short from large distances. This implies that the scalar action reads
\be 
S= \int d^4 x (-\frac{1}{2}(\partial \varphi_0)^2-\frac{1}{2}(\partial \phi)^2 - J(\varphi_0 + \phi) - V_{\rm int}(\varphi_0+\phi))
\ee
Notice that there is no mixing between the fields in the kinetic terms as the overlap $\int d^3x \phi\Box \varphi_0= \int \slashed{d}^3 p \phi(-p) p^2 \varphi_0(p)=0$ vanishes. On the other hand, the potential $V_{\rm int}$ mixes the fields. For instance a coupling like $\int d^3 x \phi^3(x) \varphi_0 (x)$ does not vanish as the double convolution $(\phi^3)(p)= \phi(p)\star (\phi(p) \star \phi(p))$ has a support which intersects $[0,L]$. The short distance field satisfies
\be 
\Box \bar \varphi_0 = J + \frac{\partial V_{\rm int}}{\partial \varphi_0}.
\ee
The action then becomes
\be 
S= \int d^4 x (-\frac{1}{2}(\partial \phi)^2+ \frac{1}{2}\bar\varphi_0 \frac{\partial V_{\rm int}}{\partial \varphi_0}(\bar\varphi_0+\phi) - \frac{1}{2}J\bar\varphi_0 - J\phi  - V_{\rm int}(\bar\varphi_0+\phi)).
\ee
At leading order in the interactions, we can solve 
\be 
\bar\varphi_0= \bar \Phi_0 + \int d^4y G(x-y) \frac{\partial V_{\rm int}}{\partial \varphi_0}(\bar\Phi_0+\phi)(y)
\ee
where 
\be 
\bar \Phi_0(x)= \int d^4x G(x-y) J(y).
\ee
Using $G(x)=G(-x)$ for the Feynman propagator, we find 
\be 
S= \int d^4 x (-\frac{1}{2}(\partial \phi)^2- \frac{1}{2}J\bar\Phi_0 - J\phi  - V_{\rm int}(\bar\Phi_0+\phi)).
\ee
This is the result we use in the main text. We can now expand in $\phi$ around the classical solution $\Bar\Phi_0$. Only the potential term has to be expanded. In particular, the linear coupling in $\phi$ comes from the first derivative of the interaction potential and reads 
\be 
{\cal J}^\lambda= -\frac{\partial V_{\rm int}(\bar\Phi_0+\phi)}{\partial \phi}\vert_{\bar\Phi_0}.
\ee
 Higher order terms are obtained similarly. 
\section{Integrals for radiation}  \label{a}
    
The Feynman integrals necessary to obtain scalar multipoles are of the type
    \begin{equation}
    \begin{aligned}
    &\int \slashed{d}^3 p \slashed{d}^3q \frac{e^{i\vec{p.r}}}{(\vec{p}^2+m^2)(\vec{q}^2+m^2)}\frac{(p^{\lambda_1}+q^{\lambda_1})...(p^{\lambda_D}+q^{\lambda_D})}{((\vec{p}+\vec{q})^2+m^2)^n} = (-i)^D\partial^{\lambda_1}...\partial^{\lambda_D}J(\lambda)\arrowvert_{\lambda=0} \\
    &\text{with}\hspace{0,5cm} J(\lambda) = \int  \slashed{d}^3 p \slashed{d}^3u\frac{e^{i\vec{p}.\vec{r}+i\vec{u}.\vec{\lambda}}}{(\vec{p}^2+m^2)((\vec{u}-\vec{p})^2+m^2)}\frac{1}{(u^2+m^2)^n} = \int_{\vec{u}} \frac{e^{i\vec{u}.\vec{\lambda}}I(u)}{(u^2+m^2)^n}.
    \end{aligned}
    \end{equation}
Then using Feynman's parametrization we obtain 
	\begin{equation}
	\begin{aligned}
	I(u) &= \int \frac{d^3\vec{p}}{(2\pi)^3} \frac{e^{i\vec{p}.\vec{r}}}{(\vec{p}^2+m^2)((\vec{u}-\vec{p})^2+m^2)} = \int \frac{d^3\vec{p}}{(2\pi)^3} \int_0^1 dx \frac{e^{i\vec{p}.\vec{r}}}{\left( x(\vec{p}^2+m^2) + (1-x)((\vec{u}-\vec{p})^2+m^2) \right)^2} \\
	& = \int \frac{d^3\vec{p}}{(2\pi)^3} \int_0^1 dx \frac{e^{i\vec{p}.\vec{r}}}{\left( \vec{p}^2 - 2(1-x)\vec{p}.\vec{u} + (1-x)\vec{u}^2 + m^2 \right)^2} \\
	& = \int \frac{d^3\vec{p}}{(2\pi)^3} \int_0^1 dx \frac{e^{i\vec{p}.\vec{r}}}{\left( (\vec{p}-(1-x)\vec{u})^2 + ((1-x)-(1-x)^2)\vec{u}^2 + m^2 \right)^2} \\
	& = \int \frac{d^3\vec{p}}{(2\pi)^3} \int_0^1 dx \frac{e^{i\vec{p}.\vec{r}}e^{i\vec{u}.\vec{r}(1-x)}}{\left( \vec{p}^2 + x(1-x)\vec{u}^2 + m^2 \right)^2}.
	\end{aligned}
	\end{equation}	
We also have
	\begin{equation}
	\begin{aligned}
	\int \frac{d^3\vec{p}}{(2\pi)^3} \frac{e^{i\vec{p}.\vec{r}}}{\left( \vec{p}^2 + A^2 \right)^2} & = \frac{2\pi}{(2\pi)^3}\int_0^{\infty}\frac{p^2dp}{ipr}\frac{(e^{ipr}-e^{-ipr})}{\left( p^2 + A^2 \right)^2} = \frac{1}{ir(2\pi)^2}\int_{-\infty}^{+\infty} = \frac{2i\pi}{ir(2\pi)^2}\frac{d}{dp}\left(\frac{pe^{ipr}}{(p+iA)^2}\right)\biggr\vert_{p=iA} \\
	& = \frac{2i\pi}{ir(2\pi)^2} \left( \frac{e^{-Ar}(1-Ar)}{-4A^2}-\frac{2iAe^{-Ar}}{2^3(iA)^3} \right) = \frac{e^{-Ar}}{8\pi A},
	\end{aligned}
	\end{equation}	    
using spherical coordinates and the residue theorem.
Denoting by $A^2=m^2+x(1-x)u^2$ we obtain
\begin{equation}
    \int \frac{d^3\vec{p}}{(2\pi)^3} \frac{e^{i\vec{p}.\vec{r}}}{\left( \vec{p}^2 + m^2+x(1-x)u^2 \right)^2} = \frac{e^{-r\sqrt{m^2+x(1-x)u^2}}}{8\pi\sqrt{m^2+x(1-x)u^2}}.
\end{equation}
So the expression for the generating function becomes
	\begin{equation}
	J(\lambda) = \frac{1}{8\pi} \int_0^1dx\int_{\vec{u}} \frac{e^{i\vec{u}.(\vec{\lambda}+(1-x)\vec{r})-r\sqrt{m^2+x(1-x)u^2}}}{(u^2+m^2)^n\sqrt{m^2+x(1-x)u^2}} = \frac{1}{8\pi} \int_0^1dx g(x).
	\end{equation}	   
We have defined 
\begin{equation}
	\begin{aligned}
	g(x) & = \int \frac{d^3\vec{u}}{(2\pi)^3} \frac{e^{i\vec{u}.(\vec{\lambda}+(1-x)\vec{r})-r\sqrt{m^2+x(1-x)u^2}}}{(u^2+m^2)^n\sqrt{m^2+x(1-x)u^2}} = \frac{2\pi}{(2\pi)^3i} \int_{-\infty}^{+\infty} \frac{udu}{|\vec{\lambda}+(1-x)\vec{r}|} \frac{e^{i\vec{u}.|\vec{\lambda}+(1-x)\vec{r}|-r\sqrt{m^2+x(1-x)u^2}}}{(u^2+m^2)^n\sqrt{m^2+x(1-x)u^2}}.
	\end{aligned}
	\end{equation}
Using parity we write this as
	\begin{equation}
	\begin{aligned}
	& = \frac{\pi}{(2\pi)^3i} \int_{-\infty}^{+\infty} \frac{udu}{|\vec{\lambda}+(1-x)\vec{r}|} \frac{(e^{i\vec{u}.|\vec{\lambda}+(1-x)\vec{r}|}-e^{-i\vec{u}.|\vec{\lambda}+(1-x)\vec{r}|})}{(u^2+m^2)^n\sqrt{m^2+x(1-x)u^2}} e^{-r\sqrt{m^2+x(1-x)u^2}}
	\end{aligned}
	\end{equation}
 which can be expanded as 
\begin{equation}
	\begin{aligned}
	& = \frac{2\pi}{(2\pi)^3} \int_{-\infty}^{+\infty} \frac{udu}{|\vec{\lambda}+(1-x)\vec{r}|} \frac{e^{-r\sqrt{m^2+x(1-x)u^2}}}{(u^2+m^2)^n\sqrt{m^2+x(1-x)u^2}} \sum_{k=0}^{\infty} (-1)^k \frac{(u|\vec{\lambda}+(1-x)\vec{r}|)^{2k+1}}{(2k+1)!} \\
& = \frac{2\pi}{(2\pi)^3} \sum_{k=0}^{\infty} (-1)^k \frac{|\vec{\lambda}+(1-x)\vec{r}|^{2k}}{(2k+1)!} \int_{-\infty}^{+\infty} du \frac{u^{2k+2}e^{-r\sqrt{m^2+x(1-x)u^2}}}{(u^2+m^2)^n\sqrt{m^2+x(1-x)u^2}} .
	\end{aligned}
\end{equation}
Now, with $v=\frac{u}{m}$, the integrand reads
	\begin{equation}
	\begin{aligned}
	K =\int_{-\infty}^{+\infty} du & \frac{u^{2k+2}e^{-r\sqrt{m^2+x(1-x)u^2}}}{(u^2+m^2)^n\sqrt{m^2+x(1-x)u^2}} = \int_{-\infty}^{+\infty} dv \frac{m^{2k+3}}{m^{2n+1}}\frac{v^{2k+2}}{(v^2+1)^n} \frac{e^{-rm\sqrt{1+x(1-x)v^2}}}{\sqrt{1+x(1-x)v^2}} .
	\end{aligned}
	\end{equation}
The divergent integral 
	\begin{equation}
	I = \int_{-\infty}^{+\infty} dv \frac{v^{2k+2}}{(v^2+1)^n} \frac{e^{-rm\sqrt{1+x(1-x)v^2}}}{\sqrt{1+x(1-x)v^2}} 
	\end{equation}
needs to be defined  properly.
Naively when $mr\rightarrow 0$, $I=\int_{-\infty}^{+\infty} dv \frac{v^{2k+2}}{(v^2+1)^n\sqrt{1+x(1-x)v^2}}$ diverges at $\infty$ when $2k+2-2n-1>0$. So for $k$ big enough it makes no sense. We must reinstate the $i\epsilon$ of the Feynman propagators $m^2\rightarrow m^2 - i\epsilon$, this implies
$ m\rightarrow\sqrt{m^2-i\epsilon}=m \left(1-\frac{i\epsilon}{2m^2} \right)=m-i\epsilon'$ with $\epsilon '=\frac{\epsilon}{2m}$. Now we have
	\begin{equation}
	I = \int_{-\infty}^{+\infty} dv \frac{v^{2k+2}}{(v^2+1)^n} \frac{e^{-r(m-i\epsilon')\sqrt{1+x(1-x)v^2}}}{\sqrt{1+x(1-x)v^2}}.
	\end{equation}
Taking $mr\rightarrow0$ but keeping $\epsilon'$ constant we obtain
\begin{equation}
	I = \int_{-\infty}^{+\infty} dv \frac{v^{2k+2}}{(v^2+1)^n} \frac{e^{i\epsilon'r\sqrt{1+x(1-x)v^2}}}{\sqrt{1+x(1-x)v^2}}
\end{equation}
This is an oscillating integral at $\infty$. We define it by going to the complex plane, $\sqrt{1+x(1-x)v^2}\simeq\sqrt{x(1-x)}v$ at $\infty$. When $\Im(v)>0$, $\Re(i\epsilon'vr)<0$ so it converges on a big circle in the upper half plane. Thus we have
	\begin{eqnarray}
	I && = \frac{2i\pi}{(n-1)!}\lim_{\epsilon'\rightarrow0}\frac{d^{n-1}}{dv^{n-1}} \left( \frac{v^{2k+2}}{(v^2+1)^n} \frac{e^{i\epsilon'r\sqrt{1+x(1-x)v^2}}}{\sqrt{1+x(1-x)v^2}} \right) \biggr\vert_{v=i}\nonumber \\ && = \frac{2i\pi}{(n-1)!}\frac{d^{n-1}}{dv^{n-1}} \left( \frac{v^{2k+2}}{(v^2+1)^n\sqrt{1+x(1-x)v^2}} \right) \biggr\vert_{v=i}.\nonumber \\
	\end{eqnarray}
	Therefore
	\begin{equation}
	\begin{aligned}
	K = \frac{2i\pi m^{2(k-n)+2}}{(n-1)!} \frac{d^{n-1}}{dv^{n-1}} \left( \frac{v^{2k+2}}{(v^2+1)^n\sqrt{1+x(1-x)v^2}} \right) \biggr\vert_{v=i} = \frac{2i\pi m^{2(k-n)+2}}{(n-1)!} A(n,k,x)
	\end{aligned}
	\end{equation}
	with 
	\begin{equation}
	A(n,k,x)=\frac{d^{n-1}}{dv^{n-1}}\left( \frac{v^{2k+2}}{(v+i)^n \sqrt{1+x(1-x)v^2}} \right) \biggr\arrowvert_{v=i}.
	\end{equation}
Collecting everything we find
	\begin{equation}
	g(x) = \frac{i}{2\pi} \sum_{k=0}^{\infty} (-1)^k \frac{|\vec{\lambda}+(1-x)\vec{r}|^{2k}}{(2k+1)!} \frac{m^{2(k-n)+2}}{(n-1)!} A(n,k,x).
	\end{equation}
And finally the generating functional
    \begin{equation}
    J(\lambda)=\frac{i}{16\pi^2} \sum_{k=0}^{\infty}(-1)^k \int_0^1dx \frac{|\vec{\lambda}+\vec{r}(1-x)|^{2k}}{(2k+1)!}\frac{m^{2(k-n)+2}}{(n-1)!} A(n,k,x).
    \end{equation}
The coefficient $A(n,k,x)$ can be extracted using Mathematica. The ones used in the main text are
    \begin{equation}
        \begin{aligned}
        & A(1,0,x)= \frac{i}{2 \sqrt{1 - (1 - x) x}}\\
        & A(2,0,x)= -\frac{i(1 - x) x}{4 (1 - (1 - x) x)^{3/2}} - \frac{i}{4 \sqrt{1 - (1 - x) x}} \\
        & A(2,1,x)= \frac{i(1 - x) x}{4 (1 - (1 - x) x)^{3/2}} + \frac{3i}{
 4 \sqrt{1 - (1 - x) x}} \\
        & A(3,1,x)= -\frac{5i(1 - x) x}{8 (1 - (1 - x) x)^{3/2}} - \frac{3i}{8\sqrt{1 - (1 - x) x}} - \frac{i}{8} \left( \frac{3 (1 - x)^2 x^2}{(1 - (1 - x) x)^{5/2}} + \frac{(1 - x) x}{(1 - (1 - x) x)^{3/2}} \right) \\
        & A(3,2,x)= -\frac{9i(1 - x) x}{8 (1 - (1 - x) x)^{3/2}} - \frac{15i}{8\sqrt{1 - (1 - x) x}} + \frac{i}{8} \left( \frac{3 (1 - x)^2 x^2}{(1 - (1 - x) x)^{5/2}} + \frac{(1 - x) x}{(1 - (1 - x) x)^{3/2}} \right)
        \end{aligned}
    \end{equation}
to obtain the monopole, dipole and quadrupole.

	\section{Integrals for tail effects} \label{c}
	
	\vspace{-0,4cm}
	
The tail effects necessitate to calculate the Feynman integrals
\begin{equation}
	J=\int \slashed{d}^3 p \slashed{d}^3q\slashed{d}^3 q'  \frac{ e^{-i\vec{\lambda}.\vec{p}-i\vec{\lambda'}.\vec{p'}} e^{-i(\vec{p}+\vec{p}'+\vec{k}).\vec{x}_{1,b}}e^{i\vec{k}.\vec{x}_{1,a}}}{((\vec{q}+\vec{p}+\vec{p'})^2+m^2)(\vec{q}^2+m^2)((\omega+i\epsilon)^2-\vec{p}^2-m^2)((-\omega+i\epsilon)^2-\vec{p'}^2-m^2)}.
\end{equation}
Using the Feynman parametrization and  writing $q=p+p'$,  $u=p-xq$, we obtain (taking $\epsilon\rightarrow0$),
	\begin{equation}
	\begin{aligned}
	& \int \slashed{d}^3 p \slashed{d}^3q \frac{e^{-i(\vec{\lambda}-\vec{\lambda'})\vec{p}-i(\vec{\lambda'}+\vec{x}_b)\vec{q}}}{((\omega+i\epsilon)^2-\vec{p}^2-m^2)((\omega-i\epsilon)^2-(q-p)^2-m^2)} \\
	& = \int_0^1 dx \int \slashed{d}^3 p \slashed{d}^3q \frac{e^{-i(\vec{\lambda}-\vec{\lambda'})\vec{p}-i(\vec{\lambda'}+\vec{x}_b)\vec{q}}}{((1-x)((\omega+i\epsilon)^2-\vec{p}^2-m^2)+x((\omega-i\epsilon)^2-(\vec{q}-\vec{p})^2-m^2))^2} \\
	& = \int_0^1 dx \int \slashed{d}^3 u \slashed{d}^3q \frac{e^{-i(\vec{\lambda}-\vec{\lambda'})(\vec{u}+x\vec{q})-i(\vec{\lambda'}+\vec{x}_b)\vec{q}}}{(u^2-\omega^2+m^2+x(1-x)q^2)^2}.
	\end{aligned}
	\end{equation}
Then similarly  writing $l=k+xq$,
	\begin{equation}
	\begin{aligned}
	\int \slashed{d}^3 k \frac{e^{-i(\vec{x}_b-\vec{x}_a)\vec{k}}}{(\vec{k}^2+m^2-i\epsilon)((\vec{k}+\vec{q})^2+m^2-i\epsilon)} & = \int_0^1dy \int \slashed{d}^3 k \frac{e^{-i(\vec{x}_b-\vec{x}_a)\vec{k}}}{((1-y)(\vec{k}^2+m^2-i\epsilon)+y((\vec{k}+\vec{q})^2+m^2-i\epsilon))^2} \\
	& = \int_0^1dy \int \slashed{d}^3 k \frac{e^{-i(\vec{x}_b-\vec{x}_a)(\vec{l}-x\vec{q})}}{(l^2+y(1-y)q^2+m^2-i\epsilon)^2}.
	\end{aligned}
	\end{equation}
Therefore we have 
	\begin{equation}
	J = \int_0^1dx\int_0^1dy \int \slashed{d}^3 u \slashed{d}^3 q \slashed{d}^3 q\frac{e^{-i(\vec{\lambda}-\vec{\lambda'})\vec{u}-i-i(\vec{x}_b-\vec{x}_a)\vec{l}}e^{-i(1-x)\vec{\lambda'}\vec{q}-ix\vec{\lambda}\vec{q}+iy(\vec{x}_b-\vec{x}_a)\vec{q}}}{(l^2+y(1-y)q^2+m^2-i\epsilon)^2(u^2-\omega^2+m^2+x(1-x)q^2)^2}.
	\end{equation}
Now, using residue theorem with $A=m^2-\omega^2+x(1-x)q^2$,
\begin{equation}
	\begin{aligned}
	\int \slashed{d}^3 u \frac{e^{-i(\vec{\lambda}-\vec{\lambda'})\vec{u}}}{(u^2+A)^2} & = \frac{1}{i(2\pi)^2|\vec{\lambda}-\vec{\lambda'}|} \int_{\mathbb{R}}du \frac{ue^{i|\vec{\lambda}-\vec{\lambda'}|u}}{(u^2+A)^2}= \frac{1}{2\pi|\vec{\lambda}-\vec{\lambda'}|}\frac{d}{du}\left(\frac{ue^{i|\vec{\lambda}-\vec{\lambda'}|u}}{(u+i\sqrt{A})^2}\right)\biggr\vert_{u=i\sqrt{A}} = \frac{e^{-|\vec{\lambda}-\vec{\lambda'}|\sqrt{A}}}{8\pi\sqrt{A}}.
\end{aligned}
	\end{equation}
Putting $B=m^2+y(1-y)q^2-i\epsilon$, we obtain
	\begin{equation}
	\int\frac{d^3l}{(2\pi)^3} \frac{e^{-i(\vec{x}_b-\vec{x}_a)\vec{l}}}{(l^2+B)^2} = \frac{e^{-|\vec{x}_b-\vec{x}_a|\sqrt{B}}}{8\pi\sqrt{B}}
	\end{equation}
and finally 
	\begin{equation}
	J = \frac{1}{(8\pi)^2}\int_0^1dx\int_0^1dy\int\frac{d^3q}{(2\pi)^3} \frac{e^{-|\vec{\lambda}-\vec{\lambda'}|\sqrt{A}}e^{-|\vec{x}_b-\vec{x}_a|\sqrt{B}}}{\sqrt{A}\sqrt{B}}e^{-i\vec{q}\vec{X}}
	\end{equation}
	with $\vec{X}=(1-x)\vec{\lambda'}+x\vec{\lambda}+(1-y)\vec{x}_b+y\vec{x}_a$.
Eventually we will take  $\lambda=\lambda'=0$ after taking derivatives of  $J$ with respect to $\lambda$ and  $\lambda'$. We will be able to set  $e^{-|\vec{\lambda}-\vec{\lambda'}|\sqrt{A}}=1$ as it does not play a role. Indeed, let us consider $\left(\frac{\partial}{\partial\lambda}\right)^n\left(\frac{\partial}{\partial\lambda'}\right)^me^{i|\vec{\lambda}-\vec{\lambda'}|\sqrt{A}}$. This is singular as $\lambda\rightarrow 0$ and $\lambda'\rightarrow 0$, so we regularise it by introducing $|\lambda-\lambda'|\rightarrow|\lambda-\lambda'+\delta|$ and taking the limit $\delta \to 0$. Then in the vanishing limit for $\lambda$ and $\lambda'$, the terms either vanish  or are propotional to $\delta^{ij}$. As the multipoles are traceless, this is also zero. So the only remaining  term are the ones where no derivatives acts on the exponential and in the limit $\lambda=\lambda'=0$ this term becomes unity.
As a result we have
	\begin{equation}
	\begin{aligned}
	J &= \frac{1}{(8\pi)^2}\int_0^1dx\int_0^1dy\int\frac{d^3q}{(2\pi)^3} \frac{e^{-|\vec{x}_b-\vec{x}_a|\sqrt{B}}e^{-i\vec{q}\vec{X}}}{\sqrt{A}\sqrt{B}}\nonumber \\ &= \frac{1}{(8\pi)^2}\int_0^1dx\int_0^1dy\int\frac{qdq}{(2\pi)^2iX} \frac{e^{-|\vec{x}_b-\vec{x}_a|\sqrt{B}}}{\sqrt{A}\sqrt{B}} \left(e^{iqX}-e^{-iqX}\right).
	\end{aligned}
	\end{equation}
	Taking the limit $m\rightarrow 0$ and $\epsilon\rightarrow 0$, we obtain (with $B=y(1-y)q^2$ and $A=\omega^2-x(1-x)q^2$)
	\begin{equation}
	\begin{aligned}
	J(\omega,\lambda,\lambda')= \frac{1}{16i(2\pi)^4}(I-I^*) \hspace{0,5cm}\text{with}\hspace{0,5cm} I=\int_0^1dx\int_0^1dy\int_0^{\infty}dq\frac{e^{-|\vec{x}_b-\vec{x}_a|\sqrt{y(1-y)}q}e^{iqX}}{X\sqrt{y(1-y)}\sqrt{x(1-x)q^2-\omega^2}}.
	\end{aligned}
	\end{equation}
	Then, using the change of variable $\tilde{q}=\frac{\sqrt{x(1-x)}}{\omega}q$,
	\begin{equation}
	\begin{aligned}
	I=\int_0^1dx\int_0^1dy\int_0^{\infty}d\tilde{q}\frac{e^{-\omega|\vec{x}_b-\vec{x}_a|\sqrt{\frac{y(1-y)}{x(1-x)}}\tilde{q}}e^{i\frac{X\omega}{\sqrt{x(1-x)}}\tilde{q}}}{X\sqrt{y(1-y)x(1-x)}\sqrt{\tilde{q}^2-1}} = \int_0^1dx\int_0^1dy\frac{e^{-\omega|\vec{x}_b-\vec{x}_a|\sqrt{\frac{y(1-y)}{x(1-x)}}\tilde{q}}e^{i\frac{X\omega}{\sqrt{x(1-x)}}\tilde{q}}}{X\sqrt{y(1-y)x(1-x)}\sqrt{\tilde{q}^2-1}}
	\end{aligned}
	\end{equation}
	with
	\begin{equation}
	\begin{aligned}
	U = \int_0^{\infty}d\tilde{q} \frac{e^{-\omega|\vec{x}_b-\vec{x}_a|\sqrt{\frac{y(1-y)}{x(1-x)}}\tilde{q}}e^{i\frac{X\omega}{\sqrt{x(1-x)}}\tilde{q}}}{\sqrt{\tilde{q}^2-1}} = \int_0^{\infty}d\tilde{q} \frac{e^{-\alpha\tilde{q}}e^{i\beta\tilde{q}}}{\sqrt{\tilde{q}^2-1}}.
	\end{aligned}
	\end{equation}
	Now, cutting in two the integral
	\begin{equation}
	\begin{aligned}
	& \int_0^1d\tilde{q} \frac{e^{-\alpha\tilde{q}+i\beta\tilde{q}}}{\sqrt{\tilde{q}^2-1}} \underset{\tilde{q}=\cos(\theta)}{=} -i \int_0^{\pi/2}d\theta e^{(i\beta-\alpha)\cos(\theta)} = K(i\beta-\alpha), \\
	& \int_1^{\infty}d\tilde{q} \frac{e^{-\alpha\tilde{q}+i\beta\tilde{q}}}{\sqrt{\tilde{q}^2-1}} \underset{\tilde{q}=\cosh(z)}{=} \int_0^{\infty}dz e^{(i\beta-\alpha)\cosh(z)} = \frac{i\pi}{2}H_0^{(1)}(\beta+i\alpha),
	\end{aligned}
	\end{equation}
	where we have defined the function $K$ and $H_0^{(1)}$.
	So finally
	\begin{equation}
	\begin{aligned}
	& \hspace{2cm} I = \int_0^1dx\int_0^1dy\frac{\left(K(i\beta-\alpha)+\frac{i\pi}{2}H_0^{(1)}(\beta+i\alpha)\right)}{X\sqrt{x(1-x)y(1-y)}}, \\
	& \text{with}\hspace{0,3cm} X=|\vec{X}|=(1-x)\vec{\lambda'}+x\vec{\lambda}+(1-y)\vec{x}_b+y\vec{x}_a \hspace{0,1cm},\hspace{0,3cm} \beta=\frac{\omega X}{\sqrt{x(1-x)}} \hspace{0,3cm}\text{and}\hspace{0,3cm} \alpha=\omega |\vec{x}_b-\vec{x}_a|\sqrt{\frac{y(1-y)}{x(1-x)}}.
	\end{aligned}
	\end{equation}
This is the  explicit expression for $J$  used in the main text.

\section{Memory integrals}
\label{app:mem}

In the main text, we find integrals of the form 
	\begin{equation}
	 \int\frac{d\omega}{2\pi}I_+(\omega) I_+(-\omega)\omega^2\ln(|\omega|)=\int\frac{d\omega}{2\pi}dtdt'I_+(t) I_+(t')\omega^2e^{i\omega(t-t')}\ln(|\omega|).
\end{equation}
This reads explicitly
\be -\int\frac{d\omega}{2\pi}dtdt'I_+(t) I_+(t')\frac{d^2}{dt'dt}\left(e^{i\omega(t-t')}\right)\ln(|\omega|) = -\int\frac{d\omega}{2\pi}dtdt'\dot{I}_+(t)\dot{I}_+(t')e^{i\omega(t-t')}\ln(|\omega|).
\ee
This needs regularising and we write in the limit $\epsilon\to 0$
	\begin{equation}
	\begin{aligned}	
	\int d\omega e^{i\omega(t-t')}\ln(|\omega|)= \left[\frac{e^{i\omega(t-t')}}{i(t-t')}\ln(\omega)\right]_{\epsilon}^{+\infty} + \left[\frac{e^{i\omega(t-t')}}{i(t-t')}\ln(-\omega)\right]^{-\epsilon}_{-\infty} - \int_{\mathbb{R}} d\omega\frac{e^{i\omega(t-t')}}{i\omega(t-t')}
	\end{aligned}
	\end{equation}
where we replace $\omega$ in the exponentials $\omega+i\delta$ for the first term  and the second $\omega-i\delta$, and we take the  Cauchy principal value of the integral. This gives finally
\begin{equation}
	\int d\omega e^{i\omega(t-t')}\ln(|\omega|)= - \int_{\mathbb{R}} d\omega\frac{e^{i\omega(t-t')}}{i\omega(t-t')}.
\end{equation}
As a result we have formally 
\be 
 {\cal F}^{-1}(\ln \vert \omega\vert)= -\frac{\theta (t)}{t}
\ee
where $\theta$ is the Heaviside distribution and ${\cal F}^{-1}$ is the inverse Fourier transform. We have similarly
\be
{\cal F}(\frac{\theta(t)}{ t})= -\ln \vert \omega \vert 
\ee
which will be used in the main text\footnote{The integrals are all formally divergent and the result of the Fourier transform of $\theta(t)/t$ is defined up to a constant as $\ln( \vert \omega\vert /\mu)$ where $\mu$ can be seen as a renormalisation scale \cite{Porto:2016pyg}. In the main text we have obtained the logarithmic terms for tail effects directly from the asymptotic expansion of the Hankel function of zeroth order. We have made a choice in the definition of the tail effect corresponding to $\mu=1/T$ where $T$ is the orbital period of the binary motion.}.
For the tail effect we obtain an integral over time of the form
	\begin{equation}
	\begin{aligned}
	& \int\frac{d\omega}{2\pi}I_+(\omega) I_+(-\omega)\omega^2\ln(|\omega|)= \int\frac{d\omega}{2\pi}dtdt'\dot{I}_+(t)\dot{I}_+(t')\frac{e^{i\omega(t-t')}}{i\omega(t-t')} =  \int\frac{d\omega}{2\pi}dtdt'\frac{\dot{I}_+(t)\dot{I}_+(t')}{(t-t')} \int_{-\infty}^{t-t'} du e^{i\omega u} \\
	& \hspace{4cm} = \int dt du \delta(u) \int_{-\infty}^{t-u} dt' \frac{\dot{I}_+(t)\dot{I}_+(t')}{(t-t')} = \int dt \dot{I}_+(t) \int_{-\infty}^{t} dt' \frac{\dot{I}_+(t')}{(t-t')}.
	\end{aligned}
	\end{equation}
This is the memory effect we discuss in the main text. Terms with a positive power of $\omega$ lead to local effects. For instance,	
	\begin{equation}
	 \int\frac{d\omega}{2\pi}I_+(\omega) I_+(-\omega)\omega= i\int dt \dot{I}_+(t)I_+(t).
	\end{equation}
There are no memory effects in this case and it will be the same for any positive power of $\omega$.	
	\section{Expansion of Hankel functions}
 \label{hankel}
We are looking for a series expansion of the Hankel function of order zero including  logarithmic terms. First of all consider the integral
    \begin{equation}
        \int_0^{\infty}d\theta e^{-x\cosh\theta}= \int_1^{\infty}\frac{dt}{t}e^{-\frac{x}{2}\left(t+\frac{1}{t}\right)}=\int_{x/2}^{\infty}\frac{du}{u}e^{-u-\frac{x^2}{y}}=\sum_{n\geq0}\frac{(-1)^nx^{2n}}{n!2^{2n}}\int_{x/2}^{\infty}\frac{du}{u^{n+1}}e^{-u}.
    \end{equation}
Let us calculate
    \begin{equation}
    I_n(x)=\int_{x/2}^\infty\frac{du}{u^{n+1}}e^u= \left[-\frac{u^{-n}}{n}e^{-u}\right]_{x/2}^{\infty} - \int _{x/2}^{\infty} \frac{du}{u^n}e^{-u} = \frac{1}{n}\left(\frac{x}{2}\right)^{-n}e^{-x/2} - \frac{1}{n}I_{n-1}.
    \end{equation}
This gives us a recursion all way  the down to    
$I_0 \sim -\ln(x/2)$ so we get  by recursion for the logarithmic terms $I_n \simeq \frac{(-1)^{n+1}}{n!}\ln(x/2)$.
Finally we can pick the logarithms in the series expansion
    \begin{equation}
     \int_0^{\infty} dx e^{-x\cosh\theta} \simeq -\sum_{n\geq0} \frac{x^{2n}}{n!}\ln(x/2) .
\end{equation}
Changing variable as $x=-iz$, we obtain for the the Hankel function
    \begin{equation}
        H_0^{(1)}(z)\simeq \frac{2i}{\pi}\sum_{n\geq0}\frac{(-1)^n z^{2n}}{(n!)^22^{2n}}\ln(-iz)
    \end{equation}
which is used in the main text.
 
	\section{Integral of unitary vectors} \label{d}	
	
	 Let us evaluate  $\int d\Omega\ n^L$ with $n^L=n^{i_1}\cdots n^{i_l}$  where $|\vec{n}|=1$.  $l$  must be even since no  tensor can be  constructed with an odd combination. We introduce the generating function
	\begin{equation}
	\begin{aligned}
	Z(j)=\int d\Omega\ e^{\vec{n}.\vec{j}} = 2\pi\int_{-1}^1 d\cos\theta e^{j\cos\theta}= 2\pi\frac{e^j-e^{-j}}{j}=4\pi\frac{\sinh j}{j}= 4\pi\left( 1+\frac{j^2}{3!} + \frac{j^4}{5!} +\cdots \right)
	\end{aligned}
	\end{equation}
	such that $\int d\Omega\ n^L = \frac{\partial}{\partial j^L}Z(j)$ with $\frac{\partial}{\partial j^L}=\frac{\partial}{\partial j^1}\cdots\frac{\partial}{\partial j^l}$. Then for $l=2p$,
	\begin{equation}
	\frac{\partial}{\partial j^L}Z(j) = \frac{4\pi2^p p!}{(2p+1)!} \left( \delta^{i_1i_2}\cdots\delta^{i_{2p-1}i_{2p}} + \rm{symmetries} \right)
	\end{equation}
	and for the symmetrised expression
	\begin{equation}
	\left( \delta^{i_1i_2}\cdots\delta^{i_{2p-1}i_{2p}} + \rm {symmetries}\right) = \frac{(2p)!}{2^p p!}\delta^{(i_1i_2}\cdots\delta^{i_{2p-1}i_{2p})} . 
	\end{equation}
	Indeed, there are $(2p)!$ ways of arranging the $i_j$, $j=1\cdots 2p$, but when they are paired, there are $2^p$ irrelevant contributions (by symmetry) and we can place the $p$ pairs in any order so $p!$. Finally we obtain \cite{Ross:2012fc}
	\begin{equation}
	\int d\Omega\ n^L = \frac{4\pi}{(2p+1)} \delta^{(i_1i_2}\cdots\delta^{i_{2p-1}i_{2p})} = \frac{4\pi}{(2p+1)!!} \left( \delta^{i_1i_2}\cdots\delta^{i_{2p-1}i_{2p}} + {\rm symmetries} \right) .
	\end{equation}

 \bibliography{ref}
 \end{document}